\documentclass[useAMS,usenatbib]{mnras}
\usepackage{graphicx}
\usepackage{epstopdf}
\usepackage{enumerate}
\usepackage{amsmath}
\usepackage{array}
\def\be{\begin{equation}}
\def\ee{\end{equation}}
\usepackage{float}

\newcommand{\sgra}{Sgr~A*}

\title[Using mass-scaling to study accretion]{Mass-scaling as a method to constrain outflows and particle acceleration from low-luminosity accreting black holes}

\author[R. M. T. Connors et al.]
    {R. M. T.~Connors,$^1$\thanks{E-mail: r.m.t.connors@uva.nl}
      S.~Markoff,$^1$ M. A. Nowak,$^2$ J. Neilsen,$^2$ C. Ceccobello,$^1$ 
      \newauthor P. Crumley,$^1$ C. S. Froning,$^3$ E. Gallo,$^4$  J. E. Nip$^{1}$\thanks{Current e-mail: jonathannip@gmail.com}\\
      \\
     $^1$Anton Pannekoek Institute, University of Amsterdam, Science Park 904, 1098 XH Amsterdam, The Netherlands\\
     $^2$Massachusetts Institute of Technology, Kavli Institute for Astrophysics, Cambridge, MA 02139, USA\\
     $^3$Department of Astronomy, C1400, University of Texas at Austin, Austin, TX 78712, USA\\
     $^4$Department of Astronomy, University of Michigan, Ann Arbor, MI 48109-1042, USA}
\date{}

\pagerange{\pageref{firstpage}--\pageref{lastpage}} \pubyear{2016}

\begin{document}

\label{firstpage}

\maketitle

%%%%%%%%%%%%%%%%%%%%%%%%%%%%%%%%%%%%
%%%%%%% ABSTRACT %%%%%%%%%%%%%%%%%%%%%%
\begin{abstract}
\\
The `fundamental plane of black hole accretion' (FP), a relation between the radio luminosities ($L_R$), X-ray luminosities ($L_X$), and masses ($M_{BH}$) of hard/quiescent state black hole binaries and low-luminosity active galactic nuclei, suggests some aspects of black hole accretion may be scale invariant. However, key questions still exist concerning the relationship between the inflow/outflow behaviour in the `classic' hard state and quiescence, which may impact this scaling. We show that the broadband spectra of A0620-00 and~\sgra~(the least luminous stellar mass/supermassive black holes on the FP) can be modelled simultaneously with a physically-motivated outflow-dominated model where the jet power and all distances are scaled by the black hole mass. We find we can explain the data of both A0620-00 and~\sgra~(in its non-thermal flaring state) in the context of two outflow-model scenarios: (1) a synchrotron-self-Compton dominated state in which the jet plasma reaches highly sub-equipartition conditions (for the magnetic field with respect to that of the radiating particles), and (2) a synchrotron dominated state in the fast-cooling regime in which particle acceleration occurs within the inner few gravitational radii of the black hole and plasma is close to equipartition. We show that it may be possible to further discriminate between models (1) and (2) through future monitoring of its submm/IR/X-ray emission, in particular via time lags between the variable emission in these bands.
\end{abstract}


%%%% INTRODUCTION %%%%%%%%%%%%%%%%%%%%
\section{Introduction}
\label{sec:intro}
Accreting black holes span an enormous range of masses, from black holes in stellar X-ray binaries (BHBs) of a few $M_{\sun}$ to active galactic nuclei (AGN) harbouring supermassive black holes (SMBH) ranging from $\sim 10^6$--$10^{10}~M_{\sun}$. The accretion physics of BHBs has been extensively studied, and their accretion evolution is well characterised by a disc instability model (see \citealt{las01} for a review). Observationally, BHBs (high mass and low mass alike) are classified via particular `states' based on their spectral and timing properties, of which there are many (e.g. \citealt{now95,gd03,rm06,bel10}), but the two longest-lived and thus `canonical' states are the so-called `hard' and `soft'. A basic underlying definition can be given to the hard and soft states of BHBs, wherein hard refers to an X-ray spectrum dominated by higher energies ($>10$ keV) and a non-thermal power-law spectrum, and soft is dominated by lower X-ray energies (2--10 keV) and a thermal blackbody spectrum. An idea currently under exploration concerns observational comparisons between AGN and BHBs that point to an identification of some AGN classifications with BHB states (e.g. \citealt{kjf06}).\\
\indent Whilst the classic thermal-blackbody thin accretion disc component provides a good model representation of the soft state (\citealt{sha}), there is not yet an agreed upon paradigm for the hard state. This latter situation is well demonstrated in the case of Cyg X-1 (a well-studied BHB with a high mass companion star), wherein multiple models with different inflow/outflow geometries (thermal/non-thermal coronae, jets) are all capable of explaining the observed hard-state X-ray spectrum \citep{now11}: spectral modelling of BHBs in the hard state is degenerate. Furthering our understanding of the accretion, ejection and radiative processes of BHBs thus calls for novel methods to strip out this spectral fitting degeneracy and determine how gravitational energy is re-distributed in the hard state. \\
\indent Observations of BHBs at both radio and X-ray wavelengths during the hard state, which we define now as sources having an Eddington scaled X-ray luminosity ($L_X$) in the range $L_X/L_{Edd}\equiv{l_X}\sim10^{-6}$--$10^{-2}$  \footnote{$L_{\mathrm{Edd}}=4{\pi}GMm_pc/\sigma_T = 1.25\times10^{38}\left(M/M_{\odot}\right)$ erg/s, where $G$ is the gravitational constant, $m_p$ is the proton mass, $c$ is the speed of light, $\sigma_T$ is the Thomson cross-section, $M$ is the black hole mass, and $M_{\odot}$ is the mass of the Sun.}, reveal a correlation between the respective luminosities, $L_{X} \propto {L_{R}}^{0.6-0.7}$ (\citealt{corb00,corb03,gfp03,corb08,corb13}, see also \citealt{mil11} and \citealt{gal14}). This scaling relation indicates a coupling between the radio/X-ray emission mechanisms during the hard state, pointing to a connection between the accretion flow and the jet---since radio emission in BHBs is identified with a steady compact jet, as directly imaged in BHBs GX 339-4, Cyg X-1, and GRS 1915+105 \citep{fen01,sti01,mil05}. This scaling relation also presents the possibility of breaking model degeneracies (discerning the dominant spectral components in hard state BHBs). \\
\indent This concept of scaling has broader implications when we compare these hard state BHBs with AGN showing similar compact jets, since a common scaling would imply the discovered correlation (and thus inflow/outflow coupling in these particular states) is independent of black hole mass. It has been shown that when one includes low-luminosity AGN with jet cores (low-luminosity AGN (LLAGN): including LINERS, FR1, and BL Lacs), the correlation extends to the so-called Fundamental Plane of Black Hole Accretion (FP), relating the X-ray luminosities, $L_X$, radio luminosities, $L_R$, and masses, $M_{BH}$ of the selected LLAGN and hard-state BHBs \citep{M03,F04,kfc06,plot12}. Efforts to derive these scaling laws begin by expressing all luminosities in terms of their dependence on mass and mass-accretion rate (expressed in mass-scaling Eddington units $\dot{m}=\dot{M}/\dot{M}_{\mathrm{Edd}}$, where $\dot{M}_{\mathrm{Edd}}=L_{\mathrm{Edd}}/\left(0.1c^2\right)$). For example, it can be shown through full calculation of the scaling relations that in order to satisfy the observed correlation, weakly accreting black holes must be radiating inefficiently ($L\propto\dot{m}^q$, where $q\approx2$) \citep{mncff03,hs03,plot12}; which includes synchrotron, inverse Compton and bremsstrahlung processes. \\
\indent There are prevalent difficulties with attempts to distinguish between these allowed radiative processes, and thus accretion models, capable of reproducing this $L\propto{\dot{m}^2}$ dependence, due primarily to the degeneracy in spectral modelling. For instance various radiatively inefficient accretion flow (RIAF) models \citep{ny94,yqn03} and outflow models \citep{ymf02,ycn05} have the inefficient ($q\approx2$) scaling predicted by the FP. In addition to broadband spectral modelling, degeneracies can be disentangled by introducing further observational data, such as X-ray variability studies \citep{vdk95,rm06}, broadband variability studies \citep{cas10,gan10,kal16}, variability comparisons of AGN and BHBs \citep{um05,mch06}, and polarisation measurements \citep{sha08,rf08}. Sgr A*, the SMBH at the Galactic centre, is a prime example of how combining these individual diagnostics leads to a better physical interpretation of the emission mechanisms \citep{bow03,gen03,ghez04,eck06,marr06,wit12,nei15,li15,dib16}. However, even with all such techniques, degeneracies still exist in the physical interpretation of the hard X-ray emission mechanisms of hard state BHBs and LLAGN. \cite{mar15} attempt a new approach in breaking the degeneracy between the SSC and synchrotron-dominated scenarios, testing the extent to which the scale invariance implied by the FP holds. They jointly model the broadband spectral energy distributions (SEDs) of two black holes on opposite ends of the mass scale (the LLAGN M81*, and the BHB V404 Cyg in a low-luminosity hard state), accreting at similar Eddington rates ($l_X\sim10^{-6}$). The same model with half of the fitted parameters at the same value (in mass-scaled units) provides a good fit to both sources, and a model in which synchrotron emission dominates the high-energy spectra provides the most reasonable fit. \\
\indent Now that a proof-of-concept has shown the method of joint spectral modelling provides physical insight into black holes across the mass scale, we want to extend the study to quiescence ($l_X < 10^{-6}$). Is quiescence a direct continuation of the hard state, or does the physics change below some accretion rate, as indicated by the increase in the X-ray power-law spectral index \citep{kon02,tom03,tom04,corb06,corb08,plot13}? \cite{plot15} model the broadband spectrum of BHB XTE J1118+480 in its quiescent state and compare to previous modelling of its hard state emission \citep{mai09}, showing that the transition from the hard to quiescent state of XTE J1118+480 may be characterised by a decrease in particle acceleration efficiency (see e.g. \citealt{mark10})---it is also interesting to note that the radio/X-ray correlation slope of XTE J1118+480 is consistent with those of other sources on the trend over 5 dex in $l_X$. Here we adopt the method presented in \cite{mar15}, fitting an outflow-dominated model (the details of which can be found in \citealt{mnw05} and \citealt{mai09}, from here on MNW05 and M09 respectively) to two black holes deep in quiescence yet on opposite ends of the mass scale; quiescent ($l_X\sim10^{-8.5}$) BHB A0620-00 and SMBH~\sgra~($l_X\sim10^{-9}$ during bright, non-thermal flares). \\
\indent In Sections \ref{sec:sgra} and \ref{sec:a0620} we give an overview of the theoretical and observational history (and therefore the source properties determined to date) of~\sgra~and A0620-00 respectively and a brief description of the data used in our modelling. In Section \ref{sec:model} we describe the model (including the updates made to the model in Sections \ref{subsec:cooling} and \ref{subsec:injection}) we apply to both sources. In Section \ref{sec:method} we describe the methodology behind fitting the broadband spectra. In sections \ref{sec:singlefits} and \ref{sec:jointfitting} we present results of individual fits to both sources, as well as the new joint fits. In section \ref{sec:conclusion} we discuss which of our model scenarios are most plausible when applied to both sources, and posit possible future observations and work.

 \section{\sgra}
 \label{sec:sgra}
\begin{figure}
\centering
\includegraphics[scale = 1.0,angle=270,width=0.45\textwidth]{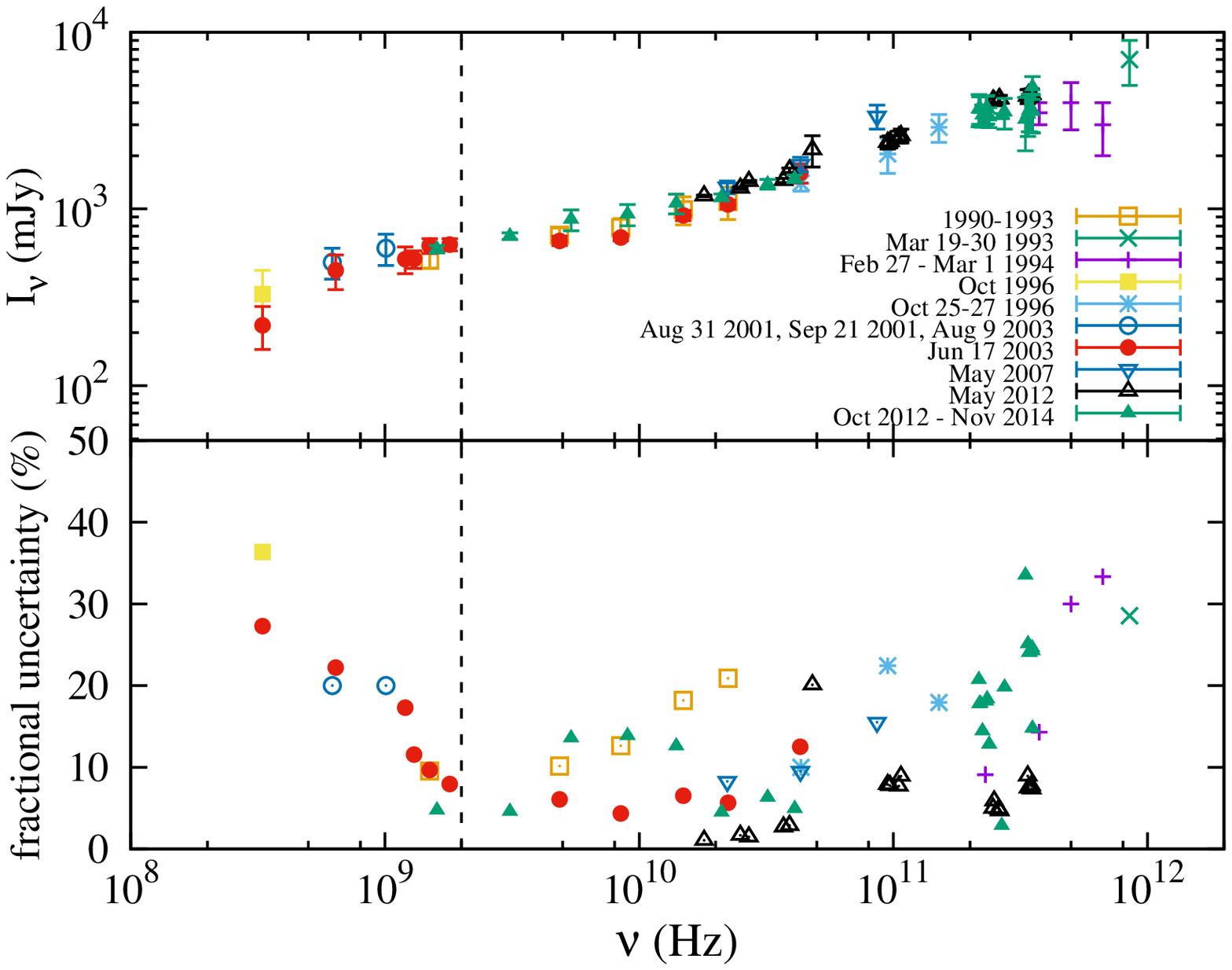}
\caption{The radio to submm spectrum of~\sgra~as observed over the past 20 years. The key shows the observing windows of the following works in order from top to bottom; \protect\cite{zha01,ser97,zyl95,nor04, fal98, rr04, an05, lu11, bri15, bow15}. The top panel shows the mean flux density as a function of frequency ranging from 330 MHz to 850 GHz. The bottom panel shows the fractional uncertainty (uncertainty/flux) of each flux measurement. The dotted line shows the boundary below which measurement uncertainty due to the scattering screen of electrons along the line of sight starts to dominate the intrinsic variability of \sgra. }
\label{fig:radio}
\end{figure}

Our own galaxy harbours an extremely weakly accreting SMBH, \sgra, that during intermittent non-thermal flaring seems to fit the criteria of a source in the universally regulated state associated with the FP (see \citealt{mf01}, \citealt{mar05}, \citealt{geg10}, \citealt{mark10} and \citealt{yn14} for full reviews on the features of~\sgra~and the Galactic centre). \sgra~has a mass of $4.1\times10^6~M_{\odot}$ \citep{gen03,ghez08,gill09} and lies at a distance of $8~\mathrm{kpc}$ \citep{rei93,ghez08,rei09}. The unabsorbed X-ray (2--10 keV) luminosity of~\sgra~during quiescence is a few times $10^{33}$ erg s$^{-1}$ \citep{bag03} or $l_X\sim10^{-11}$, making it the most weakly accreting black hole observed to date. \cite{wang13} present the results of 3 Msec of \textit{Chandra X-ray Observatory} imaging of the Galactic centre as part of an X-ray Visionary Project (see http://www.sgra-star.com), resolving the accreting gas around~\sgra~during quiescence. The results confirm that the steady quiescent spectrum of~\sgra~can be fit with a thermal bremsstrahlung model from a hot plasma near the Bondi radius, consistent with earlier predictions by e.g. \cite{nym95,quat02}. \\
\indent Frequent X-ray monitoring of~\sgra~also resulted in the discovery of flares \citep{bag01} lasting as long as 10 ks with a peak luminosity $\sim$ 50x brighter than the quiescent emission, with the flare emission best fit by a power law with a significantly harder spectrum than that detected in quiescence ($dI_{\nu}/d{\nu}={\nu}^{-\alpha}$, $\alpha\sim0.3$, with $\alpha\sim$1.2 in quiescence). Subsequent observations of the flare emission have found peak luminosities reaching 130x \citep{now12} and 400x (Haggard et al., in prep.) higher than the quiescent level, originating from much smaller radii than the quiescent emission. Attempts to model the variable (flare) emission of~\sgra~now include jet models capable of producing a synchrotron + inverse Compton (in particular SSC) \citep{fm00,mfyb01}, and hybrid models including RIAF components, both thermal \citep{ymf02} and with a non-thermal population of particles \citep{yqn03}. The 3 Msec of additional observing time with \textit{Chandra} allowed the first detailed observations of the flaring emission, doubling the population of known flares within a year \citep{nei13}. These flares range in duration from a few 100 seconds to 8 ks, and in luminosity from $\sim 10^{34}$ erg s$^{-1}$ to 2 $\times 10^{35}$ erg s$^{-1}$, bringing~\sgra~to fluxes consistent with the FP relation (in quiescence~\sgra~lies $\sim2$ orders of magnitude in $L_X$ below the FP relation). The timescales of the flares indicate an emission region of 5--400 $r_g$, though large excursions from quiescence likely originate from within $\approx5~r_g$ \citep{bar14}. \\
\indent We are interested in modelling~\sgra~during bright X-ray flares (when \sgra~approaches the FP relation), and thus we select the 3 brightest X-ray flares, with peak count rates 0.15--0.25 cts/s, whereby this grouping of flares contains sufficient cumulative counts for us to perform $\chi^2$ statistics---see Figure \ref{fig:flares}. The details of the \textit{Chandra} observations and data reduction can be found in \cite{nei13}, and in section \ref{subsec:fitmeth} we detail how the spectra are binned/grouped and subsequently modelled. \\
\indent X-ray flares observed from~\sgra~are coincident with an IR counterpart, but the opposite is not always true \citep{eck06,hor07}. This characteristic hints at the nature of the physical connection between the IR/X-ray emission, however lack of coverage combined with uncertainties regarding the IR flux distribution make simultaneous modelling a difficult task \citep{dod11,trap11,wit12}. We thus select the median IR H and ks-band fluxes found by \cite{brem11}, $3.61\pm1.62$ and $6.03\pm1.85$ mJy respectively, and the mid-IR $3\sigma$ upper limit of 58 mJy found by \cite{hau12}. Using the median NIR fluxes allows us to somewhat represent the flux uncertainties during the brightest X-ray flares, whilst the mid-IR upper limit allows us to put prior constraints on our model parameters (since the thermal synchrotron spectrum cannot exceed this upper limit). \\
\indent We require a quasi-simultaneous broadband spectrum to perform time-independent modelling. \sgra~only becomes significantly variable (up to $\sim40\%$) at submm wavelengths, when the emitting region is optically thin \citep{lu11,bow15}, though there has been a rise in flux of $\sim20\%$ in the 5--20 GHz range over the past decade \citep{an05,bow15}, as shown in Figure \ref{fig:radio}. We compile an average radio-to-submm spectrum that encompasses this short and long term variability, with appropriate coverage across the 330 MHz - 850 GHz range. The resulting data table can be found in Table \ref{tab:A1} in the Appendix.

\begin{figure}
\label{fig:flares}
\centering
\includegraphics[scale = 0.7,width=0.45\textwidth]{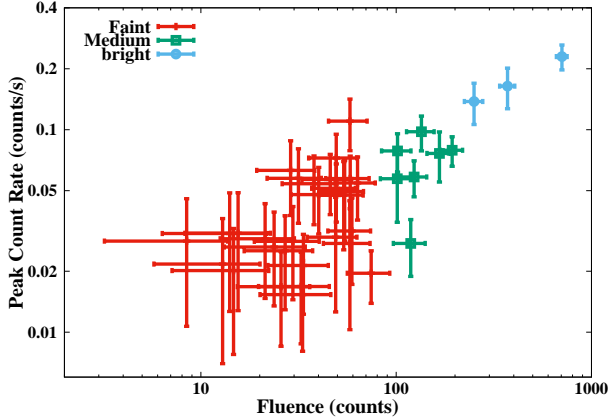}
\caption{The splitting of X-ray flares into 3 categories of peak count rate. The plot show how the 39 X-ray flares \citep{nei13} are divided into 3 sections, based on CF levels $\le$ 1200 (red lines), 1200 $\le$ CF $\le$ 2200 (green squares), and CF $>$ 2200 (blue circles).}
\label{fig:flares} % label for this figure, reference with \ref{fig:flares}
\end{figure}

 \section{A0620-00}
 \label{sec:a0620}
First discovered at X-ray wavelengths when it went into outburst in 1975 \citep{elv75}, A0620-00 (hereafter A0620) settled into quiescence 15 months later. It has been in a persistent quiescent state since then, and we now know that the system consists of a K-type donor star transferring mass to a black hole via an accretion disc \citep{mr86}. The mass, distance, and orbital inclination ($i$) are found by \cite{can10} to be $6.6\pm0.25~M_{\odot}$, $1.06\pm0.12~\mathrm{kpc}$ and $51.0^{\circ}\pm0.9$ respectively. \cite{gar01} and \cite{kon02} find a quiescent X-ray luminosity for A0620 of $3 \times 10^{30}$ erg s$^{-1}$, which corresponds to $\sim10^{-8.5}L_{\mathrm{Edd}}$, whilst \cite{gal06} find $L_X=7.1^{+3.4}_{-4.1}\times10^{30}\mathrm{erg s}^{-1}$, which also puts A0620 in the Eddington range $l_X\sim10^{-9}$--$10^{-8.5}$. Given the implication of the FP that black holes accreting at similar Eddington rates should regulate their output in the same way, A0620 is a suitable candidate for a comparison study with~\sgra~in its non-thermal `flaring' state. \\
 \indent A0620 has an 8.5 GHz radio flux density of $51\pm7~\mu{\mathrm{Jy}}$ \citep{gal06}, interpreted as self-absorbed synchrotron emission from a jet/outflow. Comparison of the radio/X-ray flux confirms A0620 as the lowest-luminosity source on the FP. Mid-IR detections suggest the self-absorbed synchrotron emission extends up to the mid-IR given the flat spectral index between radio-mid-IR, though a circumbinary disc component cannot be ruled out \citep{mm06,gal07}. \\
\indent \cite{fron11} present a broadband spectral energy distribution (SED) including X-ray, UV, optical, NIR, and radio observations of A0620, adding to the already existing broadband coverage \citep{nar96,mr00,gal06,gal07}. Through modelling of the broadband spectrum, \cite{fron11} show that $90~$\% of the disc mass is lost between the outer and inner accretion flow, indicative of either an outflow prior to capture by the black hole (an ADIOS-like solution, e.g. \citealt{bb99,bb04}), or mass loss to the black hole (an ADAF-like solution, e.g. \citealt{ny94}). \cite{wang13} find a similar accretion disc density profile explains the low accretion rates onto~\sgra, providing a further analogy between black holes of varying mass accreting at very sub-Eddington rates. Since A0620 extends the radio/X-ray correlation down to the most quiescent luminosities, and we know that brighter hard-state sources show a radio jet, it seems reasonable to assume the presence of a jet at the lowest luminosities. \\
 \indent We model the full radio-to-X-ray quasi-simultaneous spectrum of A0620. This includes the simultaneous radio/IR/optical/X-ray observations taken in August 2005 presented by \cite{gal06}, the IR observations taken 5 months prior in March 2005 by \citep{gal07}, and full IR/optical/UV observations taken by \cite{fron11} in March 2010. We refer the reader to the relevant observational papers for a full description of the data reduction and analysis. Combining these datasets gives us good coverage from radio to X-ray frequencies whilst accounting for the optical/UV variability of A0620 during its ``active" state \citep{can08}. Although this results in added $\chi^2$ residuals in our fits, it is more informative to include this variability and have a representative time-averaged spectrum. We also note that the constraints that come from fitting across 8 orders of magnitude in spectral energy outweigh the residuals accrued by modelling data over these two epochs (see e.g. \citealt{mar08}). We deredden the IR-FUV fluxes with $E(B-V)=0.39$ in agreement with \cite{gal07}. The full radio-FUV dereddened flux values are shown in table \ref{tab:A2} in the Appendix. The X-ray spectrum is identical to that modelled in \cite{gal07}.
 
 %%%% THE MODEL SECTION %%%%%%%%%%%%%%%%
 \section{The model}
 \label{sec:model}
 We explore statistical fits of the \texttt{agnjet} model (MNW05, M09) to multiwavelength spectra of both~\sgra~and A0620 separately, and then perform joint fitting of both sources, tying parameters that represent the scale invariance (this is discussed in detail in Section \ref{sec:jointfitting}). \\
 \indent MNW05 and M09 (and references therein) give a full description of \texttt{agnjet} including its assumptions and parameters, and subsequent work explores model fits to both BHBs and LLAGN \citep{plot15,mar15,pri15}. Here we give a basic outline and a description of the \texttt{agnjet} parameters. In \texttt{agnjet}, a relativistic plasma of adiabatic index $\Gamma = 4/3$ is injected in a nozzle at the base of the jet (or rather both axially symmetric jets) following assumptions for the hydrodynamics as laid out in \citealt{fb95,fal96}. At the jet base, the internal energy density, $U_J=U_B+U_e+U_{tu}$ (where $U_B$ is the magnetic energy density, $U_e$ is the relativistic electron energy density, and $U_{tu}$ is the turbulent plasma energy density), is assumed to be equal to the rest-mass energy density. \cite{z16} points out that in fact the internal energy density could be arbitrarily large. However, the small Lorentz factors found in BHB jets  ($\gamma_j\sim$ a few), as well as implied by the variability in \sgra \citep{fmb09,bri15}, require that the jet's internal energy density not exceed the rest-mass energy density by a factor of more than a few at the base of the jet. We therefore suggest that such a scenario applies for all low-luminosity sources (lying on the FP) such as~\sgra~and A0620. Limiting the internal energy density of the jet in this way means we are not describing what would be classed as a Poynting-flux dominated jet \citep{bz77} in any of our modelling; we cannot have $U_B>nm_pc^2$, since the dynamics are not correctly calculated in such a scenario. In that sense our jet is consistent with being matter-dominated \citep{bp82}.\footnote{We are currently exploring extending the \texttt{agnjet} model to cases in which the internal energy density of the plasma is not equal to the rest-mass energy density at the jet nozzle. We will address this topic in detail in a forthcoming research note (Crumley et al., in prep.), since it resides outside the scope of this paper.} The plasma is assumed to expand freely with an initial sound speed $\beta_{s,0} = \sqrt{\Gamma\left(\Gamma-1\right)/\left(\Gamma+1\right)}\sim0.43$ in the lateral direction, and longitudinal pressure gradients accelerate the jet to supersonic speeds along its axis.\\
 \begin{table}
 \centering
  \caption{A list of the main input parameters of the \texttt{agnjet} model}
  \label{tab:params}
 \begin{tabular}{@{}cp{6cm}}
 \hline
 \hline
 Parameter & Description \\
 \hline 
 $N_j$ & the normalised jet power, in units of $L_{\mathrm{Edd}}$.\\
 \hline
 $r_0$ and $h_0$ ($r_g$) & the radius and height (length) of the jet nozzle. \\
\hline
 $T_e$ (K) & the electron temperature of the input distribution. \\
 \hline
 $k$ & the ratio of magnetic to electron energy density, ${U_{B}}/{U_{e}}$, otherwise known as the partition factor. \\
 \hline
 $p$ & the power-law index of the accelerated electron distribution. \\
 \hline
 $z_{acc}$ ($r_g$) & the distance from the black hole along the jet axis where particle acceleration begins.  \\
 \hline
 $n_{nth}$ & the fraction of particles accelerated at a distance $z_{acc}$ from the black hole along the axis of the jet. \\
 \hline
 $f_{sc}$  & ${\beta_{sh}}^2/{\left(\lambda/R_{gyro}\right)}$ where $\beta_{sh}$ is the shock speed relative to the plasma, $\lambda$ is the scattering mean free path in the plasma at the shock region, and $R_{gyro}$ is the gyroradius of the particles in the magnetic field. In reality we do not require a shock so this parameterisation can generally be seen as a measure of the acceleration efficiency.\\
 \hline
 $\epsilon_{nth}$ & the fraction of energy density in non-thermal electrons injected at the jet base.\\
 \hline
 $\Delta_{fac}$ & the multiplication factor giving the maximum energy of the injected non-thermal electrons, $\gamma_{max,nth}=\Delta_{fac}\gamma_{min,nth}$.\\
 \hline
 \hline
 \end{tabular}
 \textbf{Notes}. Parameters $\epsilon_{nth}$ and $\Delta_{fac}$ only apply in model cases (c) and (d), when a mixed distribution of thermal and non-thermal particles is injected at the base of the jet.
 \end{table}
\indent The main parameters of interest are displayed in Table \ref{tab:params}. The most important fitted parameter here is $N_{j}$, the normalised jet power, since it acts as the model normalisation, and the entire spectrum is very sensitive to its value. $N_{j}$ is thus the total power fed into the base of the jet. Since the jet base represents a steady state, the inflow rate given by the power and the initial sound speed ($\beta_{s,0}$) and nozzle radius ($r_0$) sets the energy density. The partition factor $k$ parameterises the division of the jet energy density between the magnetic field and electrons ($U_B/U_e$)\footnote{This is essentially the inverse of the plasma beta parameter ($k=1/\beta$).}, where $k\sim1$ is referred to as equipartition. Since there is no radial structuring in \texttt{agnjet}, once $r_0$ is known the radial jet profile $r(z)$ is calculated, defining the jet opening angle by evaluating the velocity profile along the jet, a solution to the relativistic Euler equation for a roughly isothermal jet \citep{fal96}. The radius of the jet base $r_0$ is a very influential parameter, since the initial energy density comprising the magnetic field and particles depends inversely on the square of the radius ($U\propto{r_0}^{-2}$). Thus decreasing the jet-base radius increases the radiative output significantly, and it also affects the synchrotron/SSC radiative outputs differently. The height $h_0$ has an effect, though somewhat less than the radius. Increasing the height of the nozzle will cause an increase in the thermal synchrotron flux, as well as provide more particles to inverse Compton upscatter the synchrotron photons. We explore this parameter during fitting by both fixing it and allowing it to vary freely to explore SSC-dominated fits. \\
\indent The particles entering the base of the jet are assumed to be advected from the accretion flow with a thermal distribution at temperature $T_{e}$ (we note that this temperature may be different to the equilibrium temperature of the accretion flow, due to any heating processes at work), and are subsequently accelerated into a power law energy distribution at a distance $z_{acc}$ from the black hole along the axis of the jet (also a free parameter). An additional parameter $n_{nth}$ is included to specify the fraction of these particles accelerated into the power law (this is a separate parameterisation from $\epsilon_{nth}$, the fraction of energy density in non-thermal electrons injected at the jet base, discussed in Section \ref{subsec:injection}), which we initially fix at 0.6, reflective of typical values used in previous applications of \texttt{agnjet} (MNW05, M09). We represent the acceleration rate efficiency with the parameter $f_{sc}$, which incorporates uncertainty in the acceleration mechanism, and whose derivation is presented in \cite{jok87}. We stress that we are not asserting that the mechanism must be diffusive shock acceleration, but rather these are convenient parameterisations of acceleration in general. \\
\indent Additional parameters include the inner temperature and radius of the accretion flow, $T_{in}$ and $r_{in}$, used to describe the thermal accretion disc blackbody emission, and these photons are included in the photon field which undergoes inverse Compton scattering at low optical depth (a maximum of one scattering per photon) with the jet electrons. These are not included in Table \ref{tab:params} because for both~\sgra~and A0620 there is no discernible thin disc component to model; in the case of~\sgra~we adopt a bremsstrahlung model to fit the quiescent thermal spectrum \citep{wang13}, assumed to be produced by a RIAF as discussed in Section \ref{sec:sgra}. Fixed parameters include the mass of the black hole, $M_{BH}$, the inclination of the jet axis to the line of sight, $\theta_{i}$, and the distance to the source, $D$. \\
\indent The model \texttt{agnjet} produces 3 main components of emission---thermal synchrotron (at $z<z_{acc}$), non-thermal synchrotron (produced at $z>z_{acc}$), and SSC (with a contribution from inverse Compton scattering of disc photons, a minimal component in fits to~\sgra~and A0620)---and as previously discussed in Section \ref{sec:intro}, there is significant degeneracy between these components when fitting the spectra of hard-state BHBs and LLAGN, particularly the X-ray spectra. The jet becomes self-absorbed once it becomes optically thick at a given frequency, and hence exhibits a spectral break (also often referred to as the spectral turnover or self-absorption frequency) at $\nu_{SSA}$ below which we see the flat/inverted radio spectrum. 

% SUBSECTION - syn cooling portion %%%%%%%%%%%%%%%%%%%%%%%%%%%%%%%%%%%%%%%%%%%%%%%%%%%%
 \subsection{Synchrotron cooling}
 \label{subsec:cooling}
 The accelerated distribution of electrons had in previous versions of \texttt{agnjet} been assumed to be maintained by dissipative processes along the post-acceleration regions of the jet, with no prescription for the cooling of electrons due to synchrotron radiation (\citealt{fm00}, MNW05, M09, \citealt{plot15}, \citealt{mar15}). We have updated the code to include a prescription for synchrotron cooling that is based on solutions to the electron kinetic equation obtained by \cite{kar62}. If we consider an electron distribution in which fresh power-law electrons are continually injected and allowed to evolve with time in our adiabatically expanding jet, there will be a break in the spectrum due to the balance between supply and radiative cooling, found at 
 \be
 \label{eq:cool}
 E_{br} = \frac{4}{AB^2t} ,
 \ee
 where $A = {\sigma_{T}}/\left({6{\pi}m_e^2c^3}\right)$, $B$ is the magnetic field, $m_e$ is the electron mass. Equation \ref{eq:cool} tells us how the break energy of the electron distribution will evolve with time, but since our model is time-independent we instead quantify the break energy analytically using characteristic timescales. We do this by setting $t=t_{dyn}$ where $t_{dyn}=\Delta{z}/\beta_j{c}$ is the dynamical time during which electrons travel through a jet segment of height $\Delta{z}$ at a bulk-flow velocity $\beta_j$. Synchrotron cooling in balance with a continuous particle injection rate yields a broken powerlaw electron distribution given by 
 \be
 \label{eq:pldist}
 dN=
 \left\{
 \begin{array}{lr}
 CE^{-p}dE\,\,, & E\le{E_{br}}\\
 CE^{-\left(p+1\right)}dE\,\,, & E>{E_{br}}
 \end{array}
 \right.
 ,
 \ee
where, after substituting $t=t_{dyn}$ into (\ref{eq:cool}), $E_{br}$ is given by
\be
 E_{br}=\frac{4\beta_j{c}}{AB^2\Delta{z}}=\frac{24\pi\beta_j{m_e}^2c^4}{\sigma_TB^2\Delta{z}}.
 \ee
 
If the spectrum without cooling is given by $I_{\nu}\propto{\nu^{-\alpha}}$, then cooling produces a steepening in the spectrum from $\alpha$ to $\alpha+0.5$ at the corresponding critical break frequency, $\nu_{br}=(eB/2\pi{m_ec})\gamma_{br}^2$, where $\gamma_{br}=E_{br}/m_ec^2$. We can understand how the cooling break will evolve along the jet by considering its dependence on the variable quantities, $E_{br}\propto{\beta_jB^{-2}\Delta{z}^{-1}}$. Since the jet is accelerating, and the overall energy budget is inversely proportional to the Mach number (this is the dominant cooling term), we know that the particle number density and magnetic field strength decrease with jet height. Thus it is clear that the cooling break energy will increase with jet height, suggesting that only solutions in which acceleration occurs close to the base of the jet (preliminary fits to broadband spectra from both~\sgra~and A0620 indicate an approximate range, $z_{acc}\sim5$--$20~r_g$) will contain a synchrotron cooling break in the observed optically thin spectrum (at energies ranging from the IR and higher), assuming re-acceleration of the electrons in each zone.

% SUBSECTION - injection of mixed distribution %%%%%%%%%%%%%%%%%%%%%%%%%%%%%%%%%%%%%%%%%%%%%%%%%%%%
\subsection{Injection of a mixed particle distribution}
\label{subsec:injection}
The kinetics of the particles close to the black hole imply that the particle distribution will likely be mixed, i.e. some fraction of the particles will be non-thermal, with the bulk of the particles being thermal, and this is shown through previous modelling of \sgra's accretion flow (e.g. \citealt{yqn03,dib14}). As such we modify our model in order to allow for the possibility that a mixed particle distribution is injected at the base of the jet. The thermal particles follow a Maxwell-J\"{u}ttner distribution (as in the pure thermal case) at temperature $T_e$, where the electrons are assumed to remain relativistic ($\gamma_{min,th}=1$). Those electrons presumed to have been accelerated prior to injection are distributed as $dN=CE^{-p}dE$ between the limits $\gamma_{min,nth}=2.23~kT_e/m_ec^2$ and $\gamma_{max,nth}=\Delta_{fac}\left(2.23~kT_e/m_ec^2\right)$, where $\Delta_{fac}$ is a fixed parameter, varying only on a case-by-case basis (see Section \ref{sec:singlefits}). The fraction of electron energy density injected into the non-thermal tail is also a model parameter, $\epsilon_{nth}=U_{nth}/U_{e}$, where $U_{nth}$ is the non-thermal electron energy density, given by $U_{nth}=\int_{\gamma_{min,nth}}^{\gamma_{max,nth}}CE^{1-p}dE$. Thus $\epsilon_{nth}$ can be related to the commonly prescribed $\epsilon_e=\epsilon_{nth}/\left(1+k\right)$ which parameterises the energy given to electrons via shocks \citep{spn98}, though here we assume only that some energy is given to electrons via an unspecified acceleration process. This full distribution is then cooled both due to the jet acceleration and synchrotron emission, and we assume there is no further particle acceleration elsewhere in the jet (i.e. $z_{acc}$ becomes an inactive parameter of \texttt{agnjet}). In each zone we adjust the limits of the power-law distribution from $\gamma=1$ to $\gamma_{max,nth}$ in that zone in order to represent the thermalising of those non-thermal particles as they cool. The power-law distribution is then described by Equation \ref{eq:pldist}. In contrast to the previous case, here we simply allow the electrons to cool along the jet without re-acceleration. 

%%%%%%%%%%%%%%%%%%%%%%%%%%%%%%%%%%%%%%%%%%%%%%%%%
%%%%%%%%%%%%% METHOD %%%%%%%%%%%%%%%%%%%%%%%%%%%%%%
%%%%%%%%%%%%%%%%%%%%%%%%%%%%%%%%%%%%%%%%%%%%%%%%%
 \section{Method}
 \label{sec:method}
 We perform the spectral fits using the multiwavelength data analysis package \texttt{ISIS} \citep{hd00}, version 1.2.6-32. \texttt{agnjet} is imported into \texttt{ISIS} and (along with other model components, such as absorption routines) forward-folded through the detector response matrices. \texttt{ISIS} also allows one to read in lower frequency data (i.e. radio through to optical) from ASCII files for simultaneous broadband fitting. Any fits shown in flux space display the ``unfolded spectra," which are independent of the assumed spectral model. The model fits to the data are performed in detector space, thus all residuals are the difference between the data and forward-folded model counts, normalised by the uncertainty in that bin (standardised $\chi^2$ residuals). 
  
 % SUBSECTION - fitting methodology %%%%%%%%%%%%%%%%%%%%%%%%%%%%%%%%%%%%%%%%%%%%%%%%%%%
  \subsection{Fitting methodology}
  \label{subsec:fitmeth}
 \indent The individual fitting routines for~\sgra~and A0620 are as follows. The X-ray spectra obtained for \sgra consist of the brightest 3 of 39 total flares, as discussed in Section 4, selected based on peak count rate and total fluence (counts), ensuring we have enough photon statistics to perform our fits (see Section 5.2). The data comprise both 0th order (i.e. undispersed photons) and MEG (Medium Energy Grating) and HEG (High Energy Grating) $\pm$first order flares. Due to photon pileup 0th order spectra cannot be background subtracted, and are instead modelled as the superposition of a quiescent emission model (representative of the diffuse background emission around~\sgra; \citealt{wang13}) plus \texttt{agnjet}, with a kernel accounting for pileup. The constraints of our modelling therefore tie to the thermal quiescent spectrum. It is noted that the flares show no notable emission lines \citep{wang13}. The MEG and HEG $\pm$first order spectra are background subtracted and fit with \texttt{agnjet} corrected for interstellar absorption and dust scattering. We bin all the X-ray flare spectra at $S/N = 4$, setting a minimum number of 5 channels so as to avoid spurious groupings of photon counts at adjacent energies, and we set the energy bounds at 2--9 keV . For the 0th order and 1st orders, respectively, the fit functions are \texttt{TBnew * dustscat * (agnjet + bremss) + gaussian(1) + gaussian(2)} and \texttt{TBnew * dustscat * agnjet}, where \texttt{TBnew} represents interstellar absorption (\citealt{wam00,wilms}, with cross-sections from \citealt{ver96}), and \texttt{dustscat} accounts for dust scattering \citealt{bag03}). The \texttt{bremss} model represents the quiescent continuum associated with \sgra's accretion flow \citep{wang13}, with temperature kT $\sim$ 3.5 keV, and the two gaussian lines represent the best fit emission lines at 2.48 keV and 6.7 keV respectively, the He-like S and Fe K$\alpha$ lines (these are the strongest emission lines). We adopt $M_{BH}=4\times10^6~M_{\odot}$, $D=8~\mathrm{kpc}$, and set the jet inclination to $\theta_i=80^{\circ}$ in accordance with the orbital inclination ($0.75^{\circ}\le{i}\le{0.85}^{\circ}$) inferred from both broadband modelling \citep{mbf07} and MHD simulations of ~\sgra's accretion flow \citep{mos09,shc12,dra13}. \\ % This last sentence is dodgey!!
 \indent The X-ray spectrum of A0620 is binned at a minimum of 15 counts per bin within energy bounds 0.3--8 keV in agreement with \cite{gal06}, such that we can perform $\chi^2$ statistics on the multiwavelength spectra. Since the IR to UV spectrum is de-reddened prior to fitting, the fitting methodology is very simple. The X-ray spectra are fit with \texttt{TBnew * agnjet}, and the rest by \texttt{agnjet} alone. We are not concerned with line features present in the quiescent spectrum of A0620, however we do model the spectrum of the stellar companion, which dominates the near-IR - UV spectrum (4500 K $\le T_{star} \le$ 4900 K \citealt{gon04,fron11}). We allow the X-ray absorbing column $N_H$ to vary given the uncertainty on its measured value. We adopt $M_{BH}=6.6~M_{\odot}$, $D=1.06~\mathrm{kpc}$, and $\theta_i=51^{\circ}$ for A0620 throughout.\\
\indent For the fitting method we first make use of a fast $\chi^2$ minimisation algorithm, and then we further explore our parameter space using an \texttt{ISIS} implementation \citep{mn14} of the Markov Chain Monte Carlo (MCMC) method of \cite{fm13}. This routine makes use of the principles of an affine-invariant ensemble sampler, setting up a distribution of `walkers' in the probability density landscape. These walkers explore the landscape by accepting moves based on the probability ratio of the proposed and current positions in the parameter space. In all our MCMC runs each parameter range contains $>$ 200 walkers, all initialised uniformly within 1\% of the values found from the pre-MCMC $\chi^2$ minimisation, and allowed to evolve within flat uniform prior distributions over the area of parameter space we are exploring. Each run is allowed to evolve for at least 3000 steps in order to ensure reasonable convergence of the chain.   

%%%%%%%%%%%%%%%%%%%%%%%%%%%%%%%%%%%%%%%%%%%%%%%%%%%%%%%%%%%%%%%%%%%%%%%%%%%%%
%%%% RESULTS %%%%%%%%%%%%%%%%%%%%%%%%%%%%%%%%%%%%%%%%%%%%%%%%%%%%%%%%%%%%%%%%%%%
%%%%%%%%%%%%%%%%%%%%%%%%%%%%%%%%%%%%%%%%%%%%%%%%%%%%%%%%%%%%%%%%%%%%%%%%%%%%%
 \section{Individual spectral fits} % Individual spectral fitting results
\begin{figure*}
 % FIGURE - A0620-00 pure thermal particle injection %%%%%%%%%%%%%%%%%%%%%%%%%%%%%%%%%%%%%%%%%%%%%%%%%%%%
% \begin{figure}[h]
 \centering
\includegraphics[height=0.5\textwidth]{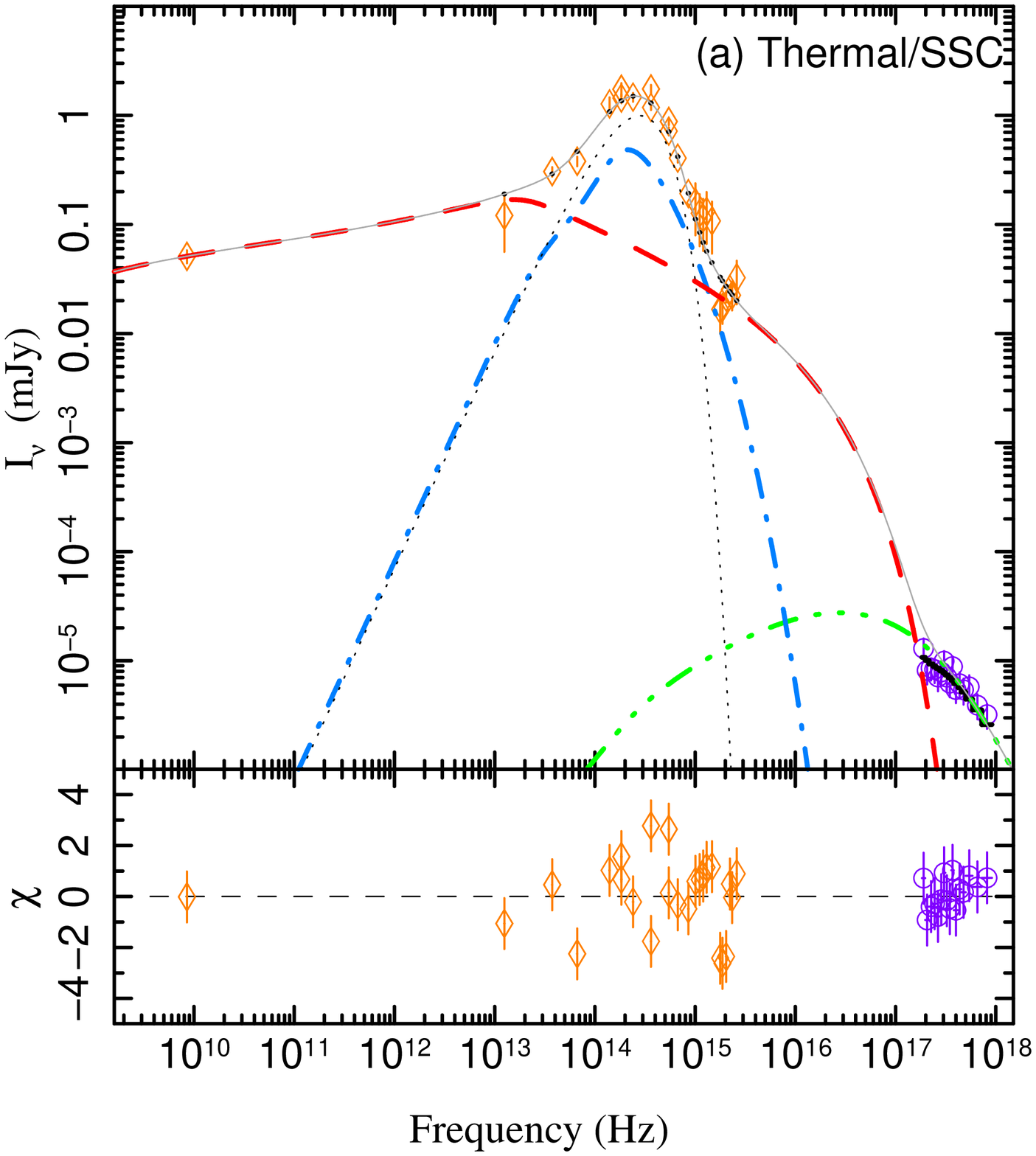}\hspace{0.4cm}\vspace{0.1cm}
\includegraphics[height=0.5\textwidth]{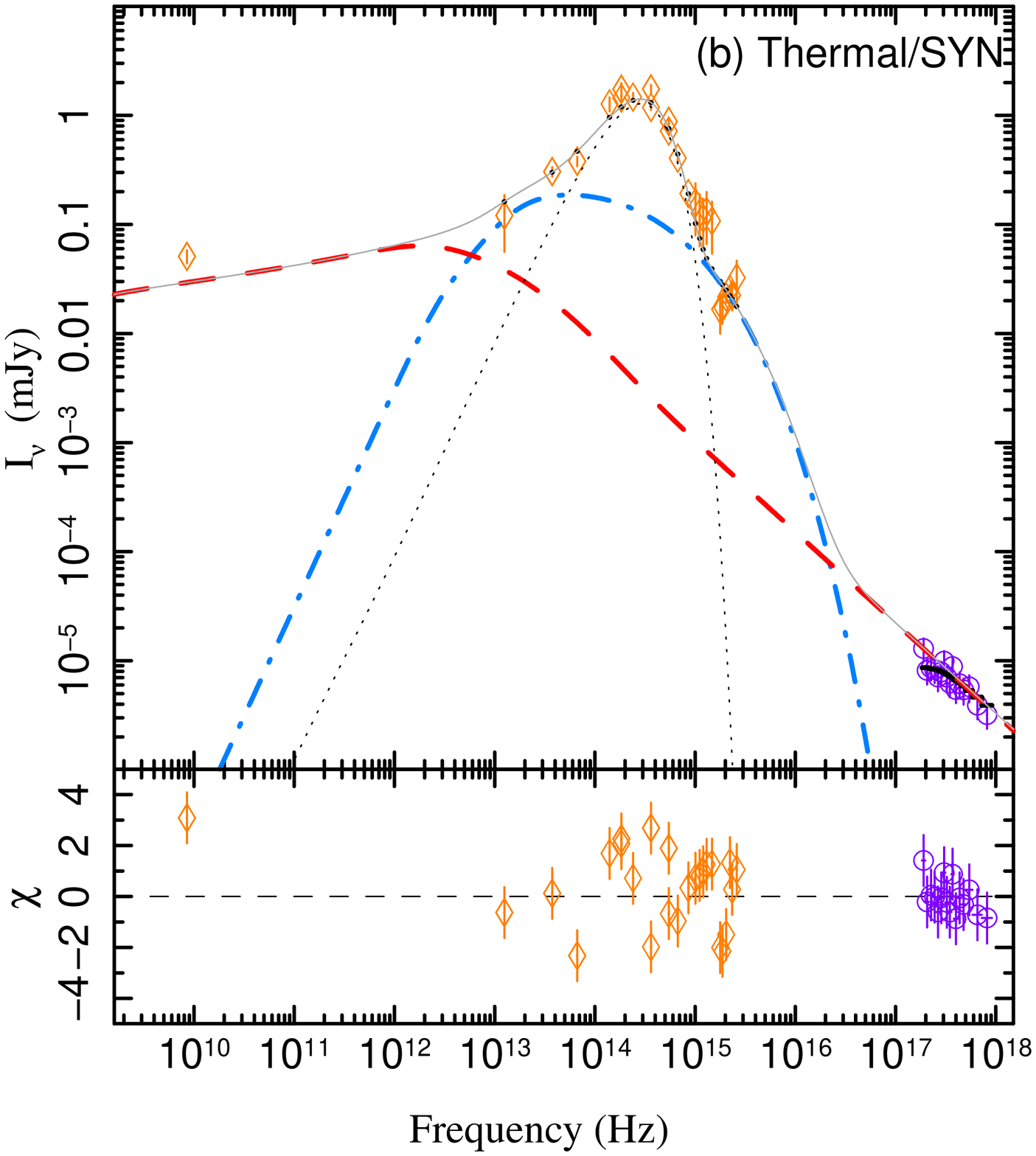}\hspace{0.4cm}\vspace{0.1cm}
\includegraphics[height=0.5\textwidth]{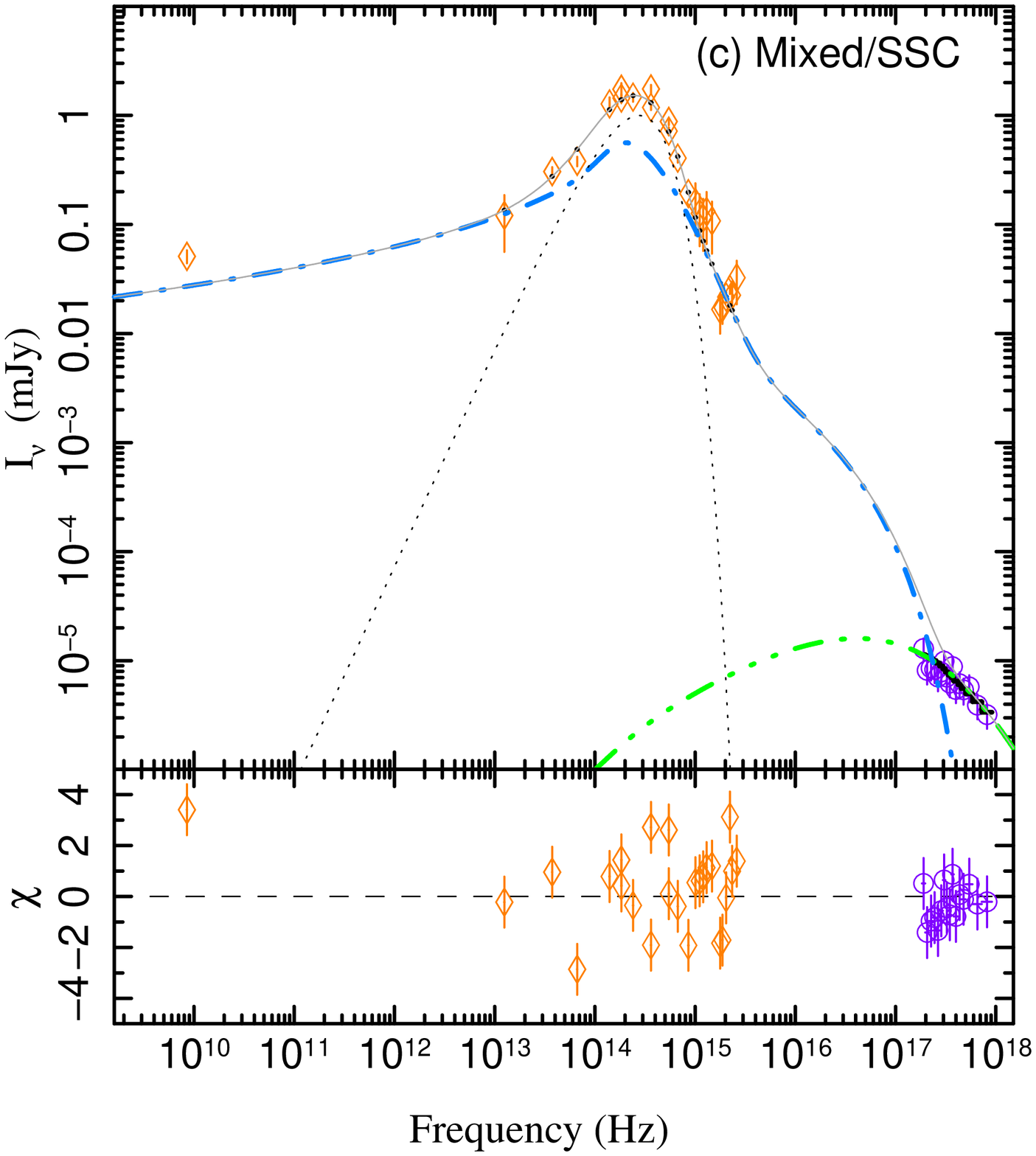}\hspace{0.4cm}\vspace{0.1cm}
\includegraphics[height=0.5\textwidth]{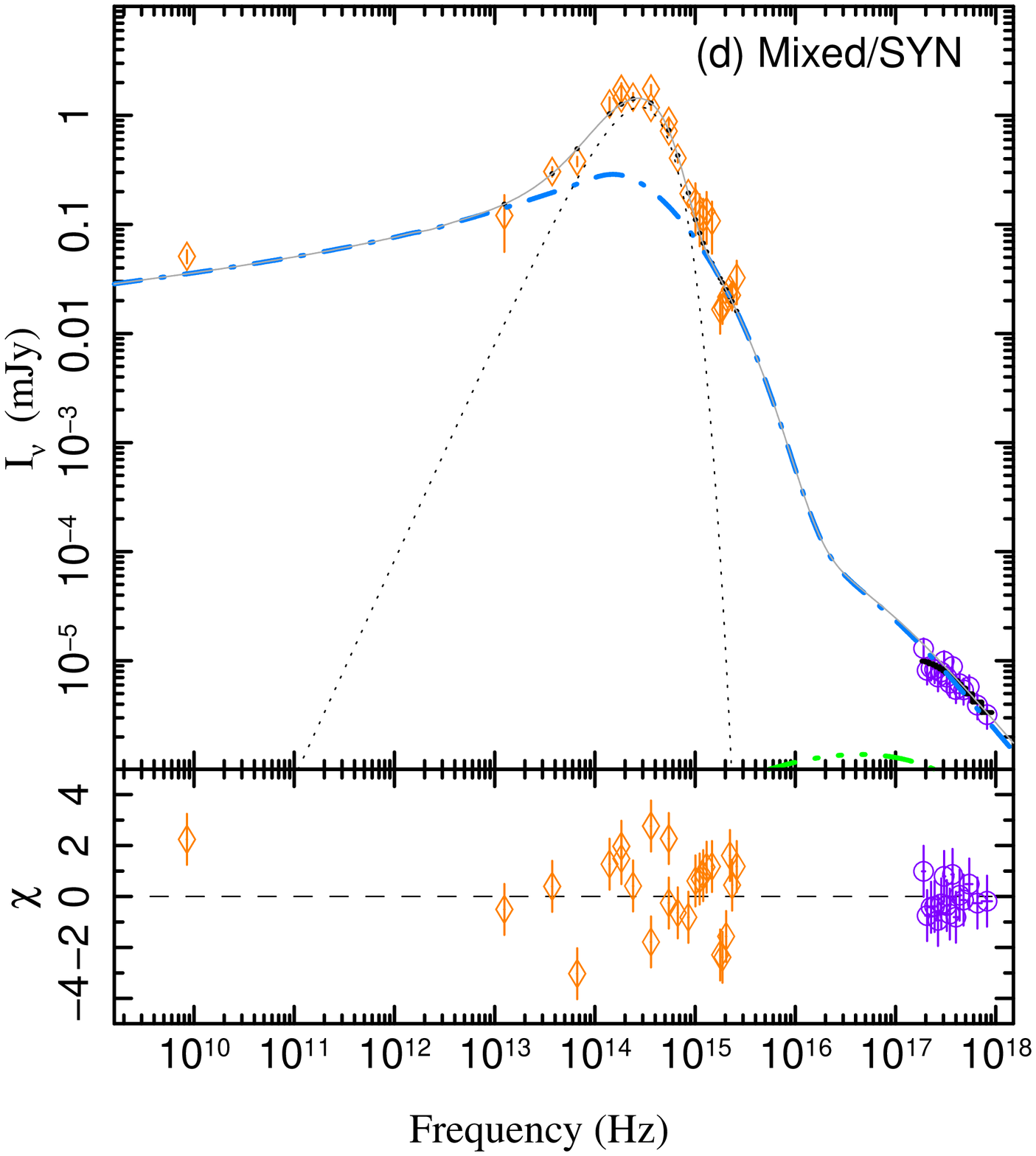}\hspace{0.4cm}\vspace{0.1cm}
 \caption{Spectral fits to the individual broadband spectrum of A0620. The four panels show cases (a) - (d) of our model fits to A0620. Orange diamond data points show the radio - FUV spectrum loaded into ISIS as flux density measurements. The X-ray spectrum is shown with purple circles. In each plot the absorbed model fit is indicated in thick black. In cases (a) and (b) the unabsorbed model components shown are pre-acceleration (thermal) synchrotron emission (blue dot-dashed line), post-acceleration synchrotron emission (red dashed line), SSC (green three-dot-dashed line), the blackbody spectrum of the stellar companion (black short-dashed line), and the total spectrum of \texttt{agnjet} (grey solid line). In cases (c) and (d) we do not include an explicit acceleration zone and thus there is no `post-acceleration' spectrum. Instead the synchrotron spectrum emitted by the full thermal + non-thermal electron distribution is shown with the blue dot-dashed line, and the SSC spectrum is shown by the green three-dot-dashed line. The bottom panels of each plot show the standardised residual photon counts. }
  \label{fig:singlefits_a0620} % figure label for a0620 fits
 \end{figure*}

\begin{figure*}
 \centering
 \includegraphics[height=0.42\textwidth]{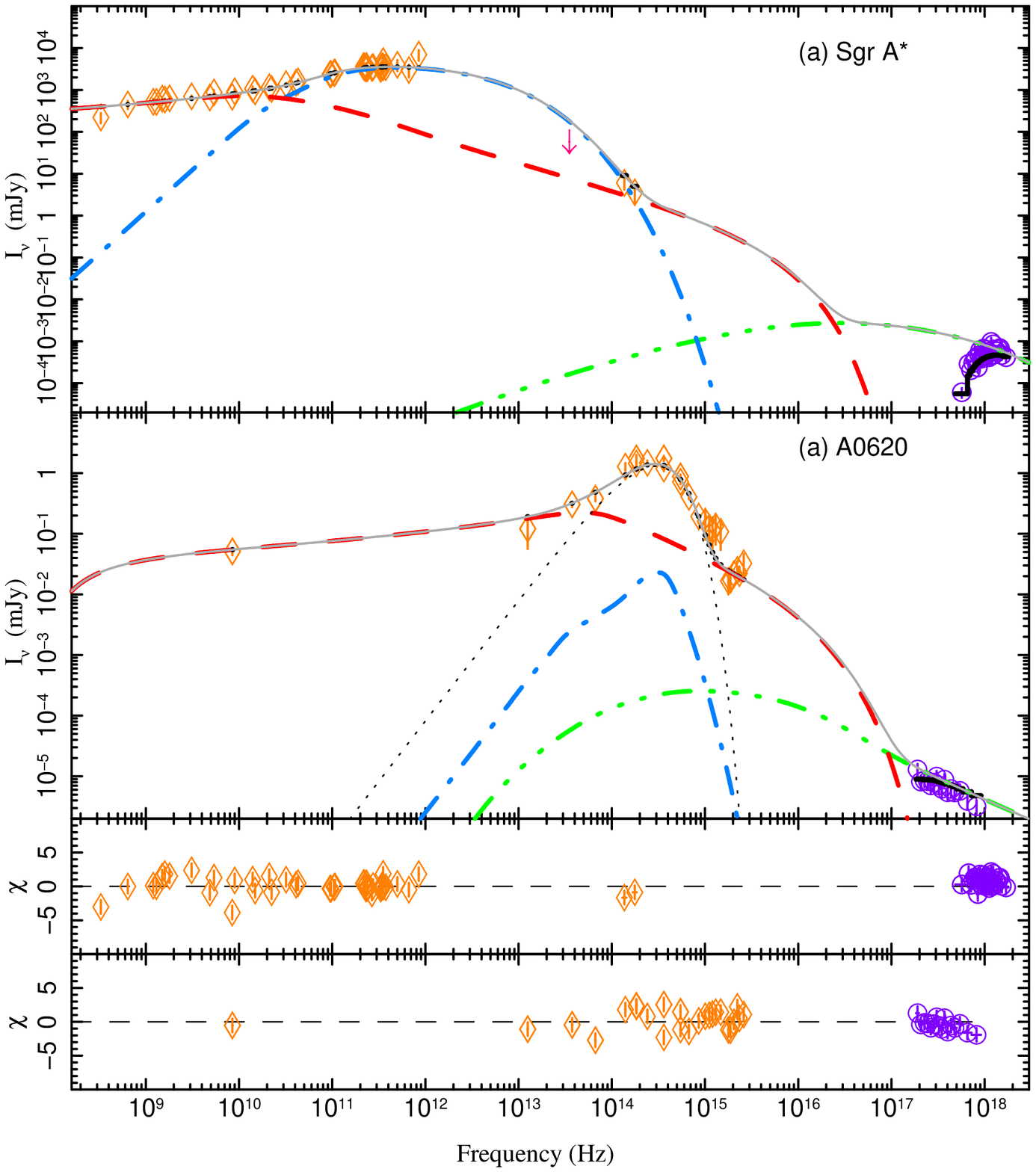}\hspace{0.4cm}\vspace{0.1cm}
\includegraphics[height=0.42\textwidth]{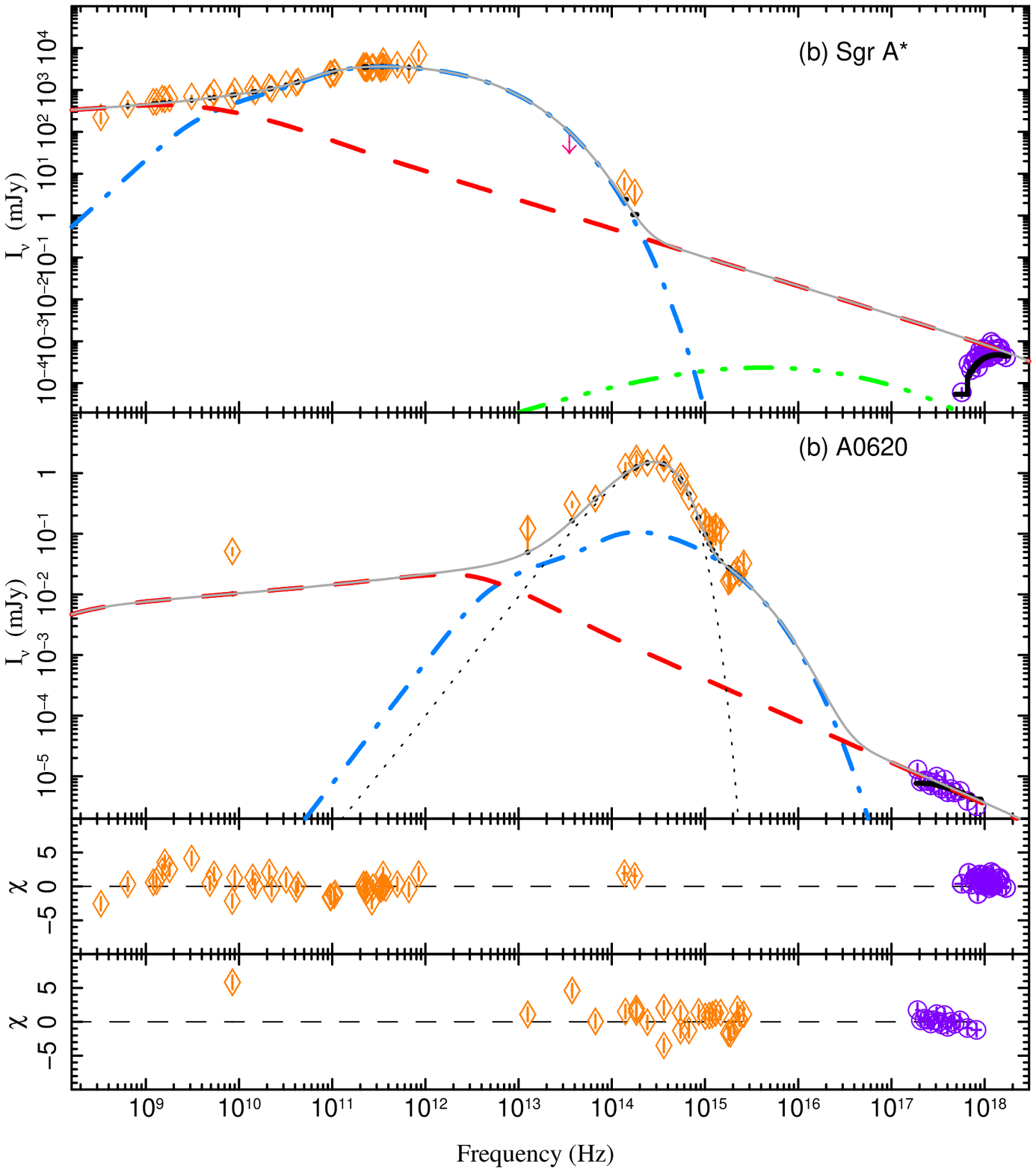}\hspace{0.4cm}\vspace{0.1cm}
\includegraphics[height=0.42\textwidth]{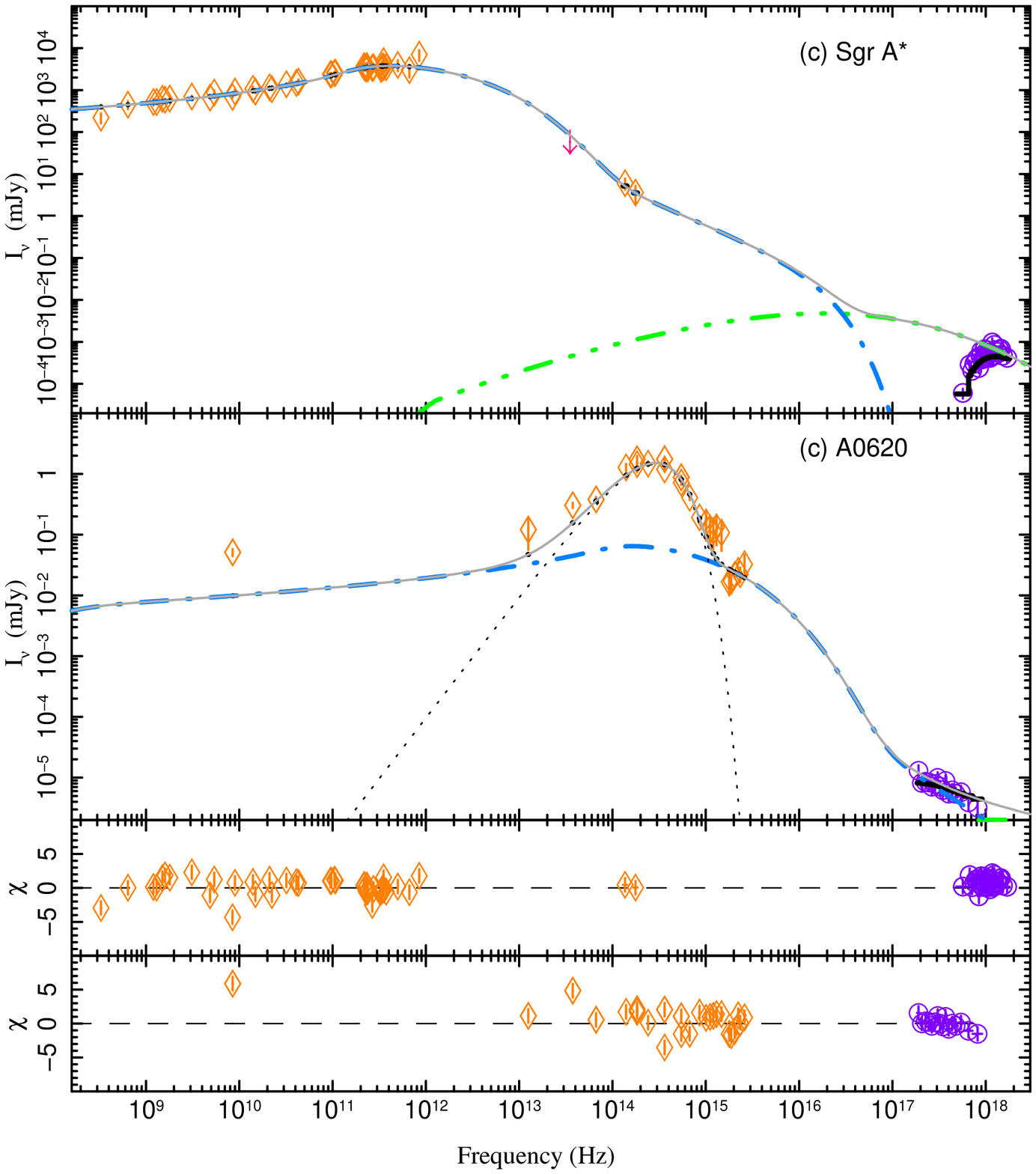}\hspace{0.4cm}\vspace{0.1cm}
\includegraphics[height=0.42\textwidth]{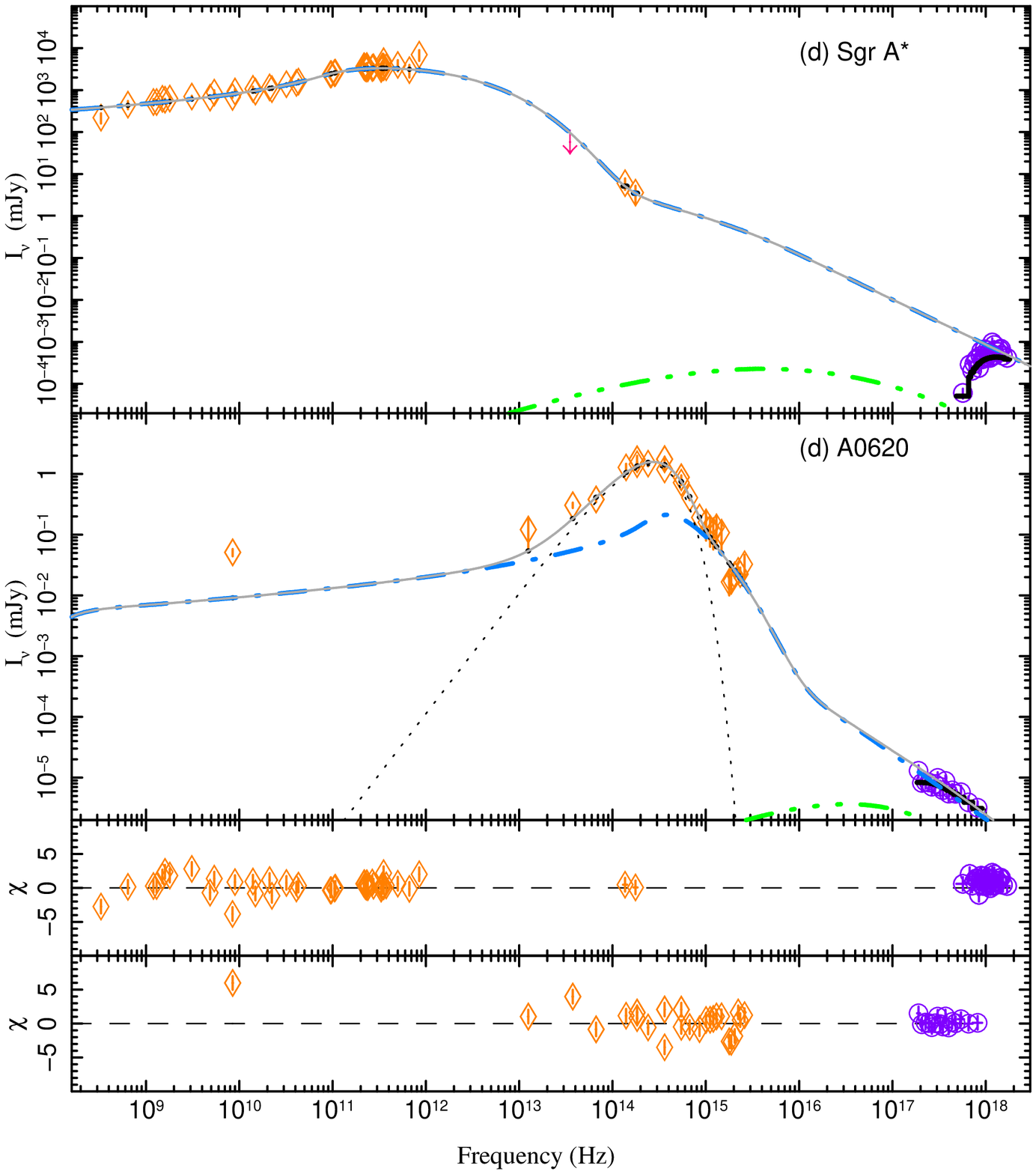}\hspace{0.4cm}\vspace{0.1cm}
\includegraphics[height=0.3\textwidth]{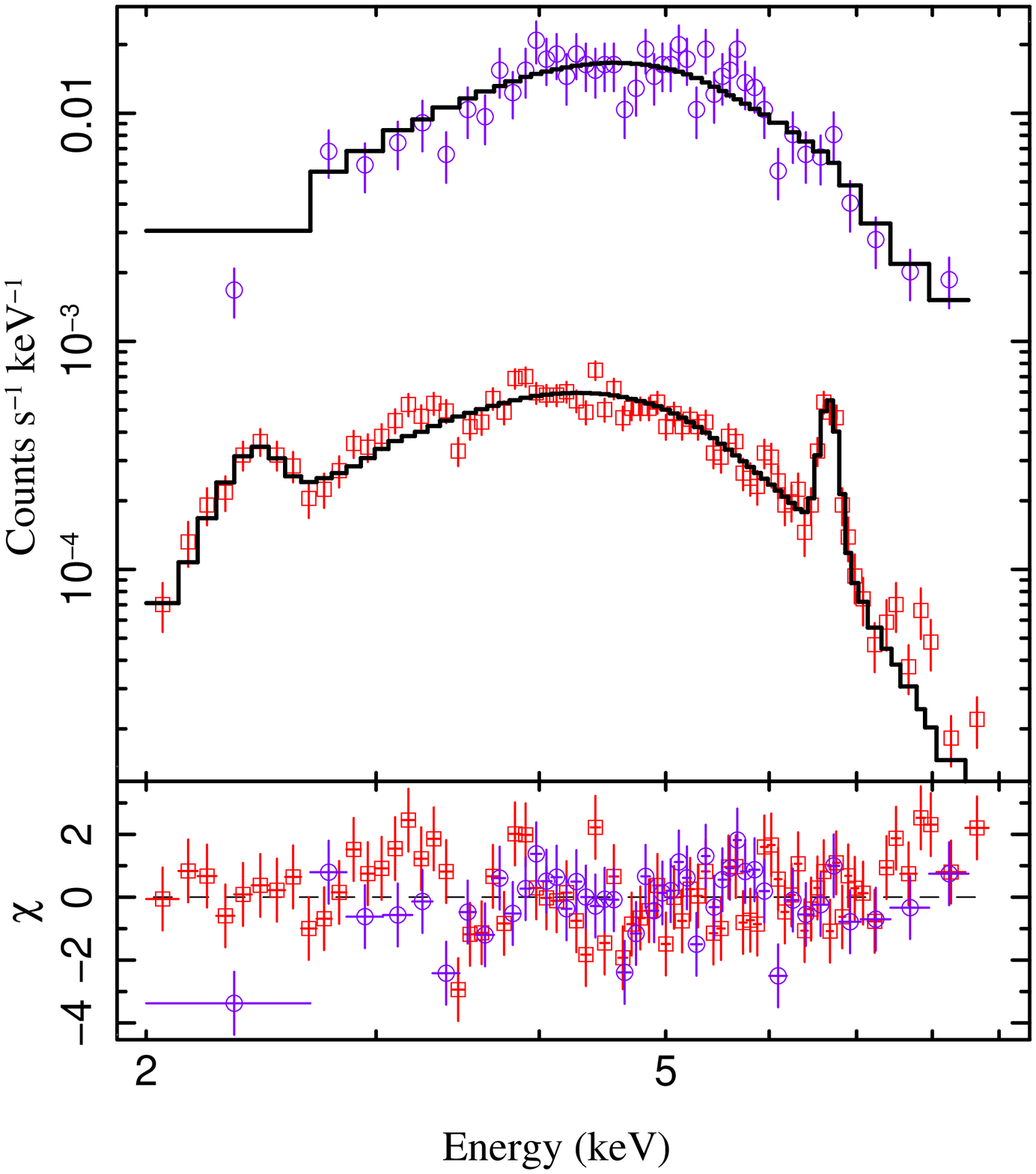}\hspace{0.01cm}\vspace{0.1cm}
 \caption{Spectral fits to the individual broadband spectrum of \sgra. The first 4 panels show cases (a) - (d) of our model fits to \sgra. Radio and IR observations are indicated by orange diamonds, and X-ray 1st-order grating spectra shown with purple circles. The mid-IR $3\sigma$ upper limit is shown as a downward pointing arrow. In each plot the absorbed model fit is indicated in thick black. In cases (a) and (b) (top panels panels) the unabsorbed model components shown are pre-acceleration (thermal) synchrotron emission (blue dot-dashed line), post-acceleration synchrotron emission (red dashed line), SSC (green three-dot-dashed line), the blackbody spectrum of the stellar companion (black short-dashed line), and the total spectrum of \texttt{agnjet} (grey solid line). In cases (c) and (d) (middle panels) we do not include an explicit acceleration zone and thus there is no `post-acceleration` spectrum. Instead the synchrotron spectrum emitted by the full thermal + non-thermal electron distribution is shown with the blue dot-dashed line, and the SSC spectrum is shown by the green three-dot-dashed line. The bottom panels of each plot show the standardised residual photon counts. The bottom panel shows the 0th-order flare- (purple circles) -and-quiescent (red squares) X-ray spectra of~\sgra~and the associated model fit, representative of all 4 fits.}
  \label{fig:singlefits_sgra} % figure label for a0620 fits
 \end{figure*}

 \label{sec:singlefits}
 Figures \ref{fig:singlefits_a0620} and \ref{fig:singlefits_sgra} show separate spectral fits to the broadband spectrum of A0620 and~\sgra~respectively, covering 4 cases for each source; (a) thermal particle injection, SSC-dominated, (b) thermal particle injection, synchrotron-dominated, (c) mixed particle injection, SSC-dominated, (d) mixed particle injection, synchrotron-dominated. The corresponding maximum likelihood estimates (MLEs) and their 90\% confidence regions are shown in Table \ref{tab:singlefits}. Here we briefly discuss the results of all 4 cases of model-fitting. Since the acceleration efficiency, $f_{sc}$, and the multiplication factor, $\Delta_{fac}$, determine the cut-off in the non-thermal particle distribution (and thus the synchrotron spectrum), we choose to fix these values ($f_{sc}$ in cases (a) and (b), $\Delta_{fac}$ in cases (c) and (d)) at their extremes. This ensures that the X-ray spectrum is fit with the non-thermal synchrotron emission in our synchrotron-dominated fits (with no cut-off present within the observing band), and the non-thermal synchrotron spectrum cuts off below X-ray energies in the SSC-dominated fits. We also note that whilst we allow the acceleration region $z_{acc}$ to be free during fitting, due to the discretised nature of the jet axis, though $z_{acc}$ is constrained it is not fully resolved; this is further complicated by possible correlations between $z_{acc}$ and the other model parameters.
  \subsection{A0620} % Concerning only the case of pure thermal particle injection
 \label{subsec:A0620_res}

 % TABLE - A0620/\sgra - list of parameters for best fits - thermal particle injection %%%%%%%%%%%%%%%%%%%%%%%%%%%%%%%%%%%
 {\small
 \begin{table*}
 \centering
 \caption{Fitted parameters for synchrotron-and-SSC-dominated individual spectral fits to A0620 and \sgra. Shown are 4 model cases for fits to each source, (a) thermal particle injection, SSC-dominated, (b) thermal particle injection,  synchrotron-dominated, (c) mixed particle injection, SSC-dominated, (d) mixed particle injection,  synchrotron-dominated. Confidence limits are at the 90\% level, a result of our MCMC exploration of the posterior distributions of the parameters. The resultant MLEs are given by the median point of each posterior distribution - this proves to be a good measure of the converged best fit of the MCMC routine. The final column shows the resultant $\chi^2$ and the degrees of freedom (DoF). From left-to-right the following parameters are shown: $N_H$, the Hydrogen column density along the line-of-sight to the source, $N_j$, the jet power, $p$, the spectral index of the power-law-distributed electrons, $T_e$, the temperature of the electron distribution (Maxwell-J\"{u}ttner), $z_{acc}$, the location of acceleration in the jet (only applicable when a pure thermal particle distribution is injected at the base, cases (a) and (b)), $r_0$, the radius of the jet-base nozzle, $h_{ratio}$ the ratio of the nozzle height $h_0$ to the jet-base radius $r_0$, $k$, the energy partition factor, and $\epsilon_{nth}$, the fraction of energy density in non-thermal electrons. Jet-base electron densities ($n_{e,0}$) and magnetic field strengths ($B_0$) are shown for the corresponding MLEs.}
 \begin{tabular*}{1.02\textwidth}{p{0.5cm}  p{1cm} p{1cm} p{1cm}  p{1.1cm} p{1cm} p{1cm} p{1cm} p{1.2cm} p{0.7cm} p{0.8cm} p{1.2cm} p{1.2cm}}
 \hline
\bf{Case} & \boldmath{$N_H$}  & \boldmath{$N_j$}  & \boldmath{$p$} & \boldmath{$T_e$}  & \boldmath{$z_{acc}$}  & \boldmath{$r_0$}  & \boldmath{$h_{ratio}$}  &  \boldmath{$k$} & \boldmath{$\epsilon_{nth}$} & \boldmath{$\frac{\chi^2}{DoF}$} & \boldmath{$\mathit{n_{e,0}}$} & \boldmath{$B_0$}\\
& {[$10^{22}$ cm$^{-2}$]} & [$10^{-7}$] & & [$10^{10}$ K] & [$r_g$] & [$r_g$] & [$h_0/r_0$]&  & [$10^{-2}$]& & [cm$^{-3}$] & [G]  \\
\\
 \hline
 \hline
 &&&&& \bf{A0620} &&&&&&& \\
 \hline
 \\
 (a)  & $0.2^{+0.3}_{-0.1}$ & $110^{+50}_{-30}$ & $1.7^{+0.5}_{-0.2}$ & $2.5^{+0.8}_{-0.4}$ & $60^{+1270}_{-30}$ & $6^{+1}_{-1}$ & $2^{+2}_{-1}$ & $0.2^{+0.3}_{-0.1}$ & ... & 58/31 & $\mathit{1.2\times10^{14}}$ & $\mathit{8.7\times10^{5}}$\\
 \\
 \hline
 \\
(b) & $0.15^{+0.10}_{-0.04}$ & $110^{+20}_{-40}$ & $2.7^{+0.3}_{-0.2}$ & $8.5^{+0.5}_{-4.9}$ & $200^{+260}_{-150}$ & $23^{+8}_{-13}$ & $1.5^f$ & $7^{+2}_{-6}$ & ... & 70/32 & $\mathit{4.0\times10^{11}}$ & $\mathit{5.1\times10^4}$\\
\\
\hline
 \\
(c) & $0.4^{+0.4}_{-0.3}$ & $210^{+110}_{-60}$ & $2^{+1}_{-1}$ & $3.3^{+1.3}_{-1.0}$ &  ... & $10^{+3}_{-2}$ & $1.1^{+0.4}_{-0.4}$ & $0.20^{+0.16}_{-0.09}$  & $<23$ & 77/31 & $\mathit{7.9\times10^{13}}$ & $\mathit{7.3\times10^4}$\\
\\
\hline
 \\
(d) & $0.12^{+0.21}_{-0.10}$ & $130^{+70}_{-40}$ & $2.2^{+0.7}_{-0.5}$ & $4.2^{+20.0}_{-0.9}$ & ... &  $9^{+85}_{-1}$ & $0.6^{+0.8}_{-0.3}$  & $2^{+7}_{-1}$ & $0.4^{+4.8}_{-0.3}$  & 67/31 & $\mathit{1.8\times10^{13}}$ & $\mathit{1.3\times10^5}$\\
\\
 \hline
 \hline
  &&&&& \bf{\sgra} &&&&&&& \\
  \hline
  \\
  (a) & $12.5^{+0.8}_{-0.7}$ & $13^{+3}_{-1}$ & $2.0^{+0.6}_{-0.3}$  & $48^{+1}_{-2}$ & $50^{+40}_{-30}$ & $2.2^{+0.2}_{-0.1}$ & $1.2^{+0.1}_{-0.1}$ & $0.003^{+0.001}_{-0.001}$ & ... & 239/181 & $\mathit{1.3\times10^7}$ & $\mathit{13.7}$\\
  \\
  \hline
  \\
  (b) & $12.9^{+0.3}_{-1.4}$ & $4.59^{+0.09}_{-0.95}$ & $2.42^{+0.02}_{-0.15}$  & $20.9^{+0.1}_{-0.9}$ & $100^{+1420}_{-20}$ & $3.9^{+0.4}_{-0.5}$ & $1.5^f$ & $9.5^{+0.4}_{-2.9}$ & ... & 274/182 & $\mathit{3.1\times10^5}$ & $\mathit{80.3}$\\
  \\
  \hline
  \\
  (c) & $12.5^{+0.8}_{-0.8}$ & $17^{+5}_{-3}$ & $2.3^{+0.5}_{-0.6}$  & $40.8^{+0.2}_{-0.6}$ & $...$ & $2.02^{+0.05}_{-0.02}$ & $0.85^{+0.05}_{-0.03}$ & $0.002^{+0.001}_{-0.001}$ & $3^{+4}_{-2}$ & 257/181 & $\mathit{2.4\times10^7}$ & $\mathit{14.4}$\\
  \\
  \hline
  \\
  (d) & $13.2^{+0.9}_{-0.7}$ & $4.6^{+0.1}_{-0.2}$ & $2.00^{+0.1}_{-0.1}$  & $18.95^{+0.05}_{-0.18}$ & $...$ & $4.1^{+0.40}_{-0.2}$ & $1.5^f$ & $9.7^{+0.3}_{-0.8}$ & $0.8^{+0.9}_{-0.4}$ & 327/182 & $\mathit{3.0\times10^5}$ & $\mathit{76.1}$\\
  \\
  \hline
 \end{tabular*}
 \label{tab:singlefits}\\
 \textbf{Notes.} $^f$ Frozen parameter
 \end{table*} 
 }

 %
 % SUBSUBSECTION - A0620-00 - fits with pure thermal distribution %%%%%%%%%%%%%%%%%%%%%%%%%%%%%%%%%%%%%%%%%%%
 %
 \subsubsection{Case (a): thermal particle injection, SSC-dominated}
 \label{subsubsec:a0620_case_a}
 This case corresponds to pure thermal particle injection at the jet base, with the X-ray spectrum dominated by SSC emission from the electrons in the base of the jet. The physical state portrayed is close to that found by \cite{gal07} in which an older version of \texttt{agnjet} is fit to a multiwavelength spectrum of A0620 (the spectral coverage of the data in their modelling was the same, but there were fewer data points in the optical/UV bands). The jet base is relatively compact, and the magnetic energy density is sub-equipartition with respect to the electrons. Such a model class coincides with those previously found to work well for BHBs in quiescence at low luminosities ($l_X\sim10^{-9}$--$10^{-8}$; \citealt{plot15}). Another distinct property we notice in this model fit when compared to cases (b), (c) and (d), is that the radio and X-ray spectra are fit simultaneously with ease. 
 \subsubsection{Case (b): thermal particle injection, synchrotron-dominated}
 \label{subsubsec:a0620_case_b}
 The synchrotron-dominated fit shows broadly different physical specifications, with higher electron temperatures than those seen in case (a), a sightly less compact jet base, and again a roughly equipartition magnetic field with respect to the electrons. No cooling break exists within the limits of the non-thermal particle distribution. This is because in all synchrotron-dominated fits in which particle acceleration occurs only at $z_{acc}$, the fits evolve to solutions in which the acceleration region is too high for efficient cooling to occur ($B_{z_{acc}}\sim0.14B_0$, where $B_0$ is the jet-base magnetic field strength). 
 
 \subsubsection{Case (c): mixed particle injection, SSC-dominated}
 \label{subsubsec:a0620_case_c}
 Here again the X-ray spectrum is SSC-dominated, except the injected distribution carries a fraction $\epsilon_{nth}$ of non-thermal energy. This fraction is poorly constrained here since it influences the synchrotron emission more than the SSC emission, but it must nonetheless be a small fraction ($<23$\%). As in case (a) the electrons are constrained to fairly low temperature and the jet base is compact and well constrained. The magnetic field is sub-equipartition, and the jet power is high and statistically distinguishable from the ranges found for cases (a) and (b), albeit not well constrained. It is not obvious that this is a physical difference, it is more likely that the method by which we divide energy between thermal and non-thermal particles in each case causes a systematic change to the injected power. 
  
 \subsubsection{Case (d): mixed particle injection, synchrotron-dominated}
 \label{subsubsec:a0620_case_d}
 If again we assume a small fraction of the particles present at the jet base are non-thermal, we can explain the X-ray spectrum with synchrotron emission from those non-thermal particles, provided non-thermal particles carry just a small fraction of the total particle energy ($<5.2$ \%, see Table \ref{tab:singlefits}). The particles must however have been accelerated quite efficiently, with their cut-off extending to $10^4~\gamma_{min}$, and a cooling break between UV/X-ray energies at $\sim10^{17}$ Hz. Again we find the jet power must be systematically higher when we inject this mixed distribution of particles in comparison to the pure thermal case, which again may simply be due to how we divide the energy between the particles. There is an increase in the upper limit on the electron temperature when compared with the SSC-dominated fit (inspection of the probability distribution reveals a bi-modal behaviour), and the system also tends towards a slightly super-equipartition magnetic field with respect to the electrons.  
 % SUBSUBSECTION -~\sgra~- fits with pure thermal distribution %%%%%%%%%%%%%%%%%%%%%%%%%%%%%%%%%%%%%%%%%%%%
 %%%%%%%%%%%%
 \subsection{\sgra}
 \label{subsec:sgra_res}
  In all fits to~\sgra~the mid-IR upper limit ($I_{8.6\mu{m}}<58$mJy, \citealt{hau12}) sets a rough flat prior on the upper limits of $T_e$ and $k$. In a mildly sub-equipartition regime ($k\sim0.1$--$0.5$) we find an approximate upper limit on the electron temperature of $T_e\sim3\times10^{11}~$K (and poor fits to the X-ray flare spectrum). Thus, in this regime, the brightest X-ray flares detected around~\sgra~cannot be explained by SSC emission as long as the plasma remains in equilibrium (see however \citealt{dib14}). To achieve higher temperatures and thus satisfactorily model the X-ray flares, we require a much more sub-equipartition flow ($k\sim10^{-3}$). Figure \ref{fig:singlefits_sgra} (panels (a) and (c)) shows SSC-dominated fits in which the jet energy partition has fallen to values of $k\sim10^{-3}$), and in this case, we find that an electron temperature of $T_e>4\times10^{11}~$K provides a good fit to the full spectrum without violating the mid-IR upper limits.
  
 \subsubsection{Case (a): thermal particle injection, SSC-dominated}
Here we require an electron temperature $T_e > 4.6\times10^{11}~$K, with a highly sub-equipartition magnetic energy density in order to model the X-ray flare spectrum of~\sgra. The size of the acceleration region is roughly consistent with the flare timescales, though too large when considering the brightest flares; \citealt{nei13,bar14}). The NIR fluxes weakly constrain the electron power law index (producing non-thermal synchrotron emission, consistent with those found by \cite{brem11}: $\alpha\sim0.7\pm0.4$, thus $p\sim2.4\pm0.8$. We do not see any strong correlation between $T_e$ and $k$ in this local fit landscape, implying that once we go to very sub-equipartition conditions, the electron temperature must always be high. 

 \subsubsection{Case (b): thermal particle injection, synchrotron-dominated}
 In this case we are able to model the X-ray flare spectrum with uncooled non-thermal synchrotron emission, in which the magnetic field is in rough equipartition with the electrons. Particle acceleration occurs at a region ($z_{acc}$) that is roughly consistent with the range of flare timescales, though again not consistent with the timescales of brightest flares \citep{bar14}; solutions like this in which particle acceleration occurs further out in the jet are thus unlikely. 
 
 \subsubsection{Case (c): mixed particle injection, SSC-dominated}
 Assuming we have a fraction $\epsilon_{nth}$ of non-thermal energy in non-thermal electrons in our jet also allows us to successfully model the~\sgra~X-ray flare spectrum in the SSC-dominated case (given a highly sub-equipartition magnetic energy density), though it should be noted that the thermal synchrotron spectrum reaches the mid-IR upper limit. The NIR spectrum is fit partially with the thermal synchrotron turnover, but is dominated by synchrotron emission from the non-thermal tail of electrons. We notice that the fit also tends towards a more compact jet base and a high jet power, which is an attempt to boost the SSC flux to match the X-ray spectrum (as inverse Comptonisation depends strongly on the electron density), though as mentioned in Section \ref{subsec:A0620_res}, the systematic increase in power when injecting a mixed particle distribution may be a natural consequence of how we divide energy between the particles. 
 
\subsubsection{Case (d): mixed particle injection, synchrotron-dominated}
A scenario in which particle acceleration occurs within a few $r_g$ of the black hole favours the flare timescales of \sgra, and such solutions are preferred in other more detailed modelling of the accretion flow close to the black hole \citep{yqn03,dib14}. Our synchrotron-dominated fit within this scenario explains both the IR and X-ray flare spectra, with slightly super-equipartition conditions at the jet base. A small fraction of the electron energy is in non-thermal electrons, producing a hard non-thermal synchrotron tail that fits the IR emission, and the X-ray spectrum is then well modelled by synchrotron emission from the cooled electrons, with a break at $\sim10^{16}$ Hz. It should however be noted that the thermal synchrotron flux sits at the mid-IR upper limit.
 
  \section{Joint Fitting}
 \label{sec:jointfitting}

 As seen in the individual spectral fits (Section \ref{sec:singlefits}), the ranges of the potential scaling parameters (which we shall now discuss) are comparable for all models, which leads well into exploring joint fits with these parameters tied. The technique used when fitting \texttt{agnjet} jointly to A0620 and~\sgra~follows the same logic as with individual fits. Each data set is loaded into \texttt{ISIS}, and the model definitions are split as described in Section \ref{sec:singlefits}, except we now require almost the same model for both~\sgra~and A0620.  \\
 \indent Following the theoretical prescription of \cite{mncff03} and \cite{hs03}, we assume that scale invariance manifests itself in geometric quantities such as $r_0$, $h_0$, and $z_{acc}$, and so we choose to tie these together for both sources. We tie a further 2 parameters by presuming that the division of energy and the acceleration mechanisms are roughly coincident at similar accretion rate in the sources that fit on the FP ($k$ and $p$); both these parameters are key to understanding whether there are fundamental differences in the energy partition of sources at very quiescent levels, and also whether spectral indices may differ \citep{rus13,plot15}. By tying these 5 parameters together, we both explore the extent to which black holes can be treated as scale-invariant, as well as potentially breaking some of the model degeneracies. Parameters that we expect to depend explicitly on cooling and the physical values of the electron density and magnetic field (e.g. $T_e$) would not be expected to scale with black hole mass, and are thus left to vary accordingly. We effectively equate the $f_{sc}$ parameter for both sources by freezing its value at its extremes in the SSC/synchrotron-dominated cases as described in Section \ref{sec:singlefits}. We refer the reader to Table \ref{tab:jointfits} for all parameter values mentioned in the following presentation of the results. We now discuss the 4 cases of joint fits just as in Section \ref{sec:singlefits}, that cannot be ruled out relative to one another in terms of their goodness-of-fit, but we discuss which areas of parameter space are less favourable given what we know about the behaviour of~\sgra~and A0620.
 
 % FIGURE - Joint fits - thermal distribution injection
 \begin{figure*}
 \centering
\includegraphics[height=0.54\textwidth]{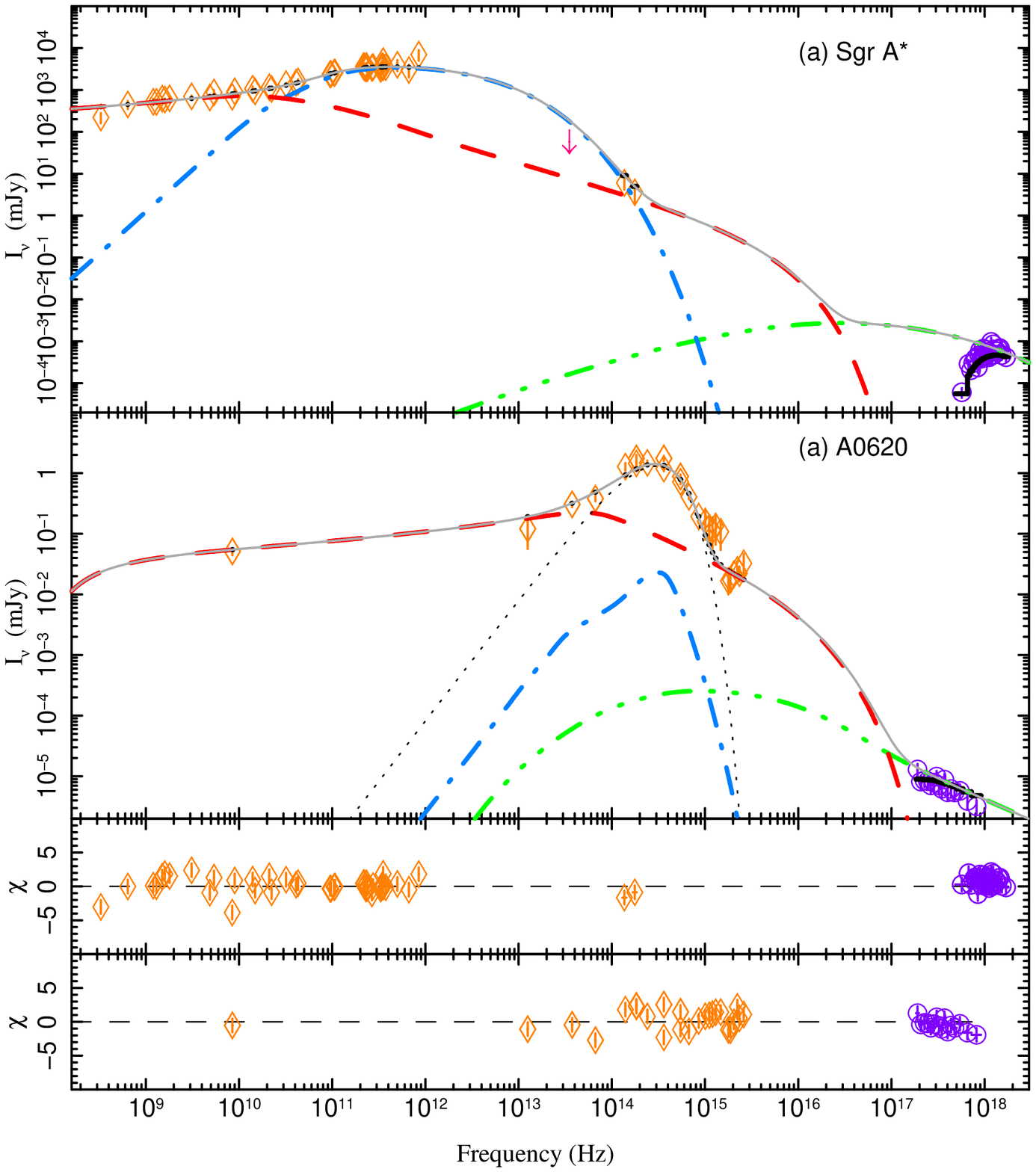}\hspace{0.4cm}\vspace{0.1cm}
\includegraphics[height=0.54\textwidth]{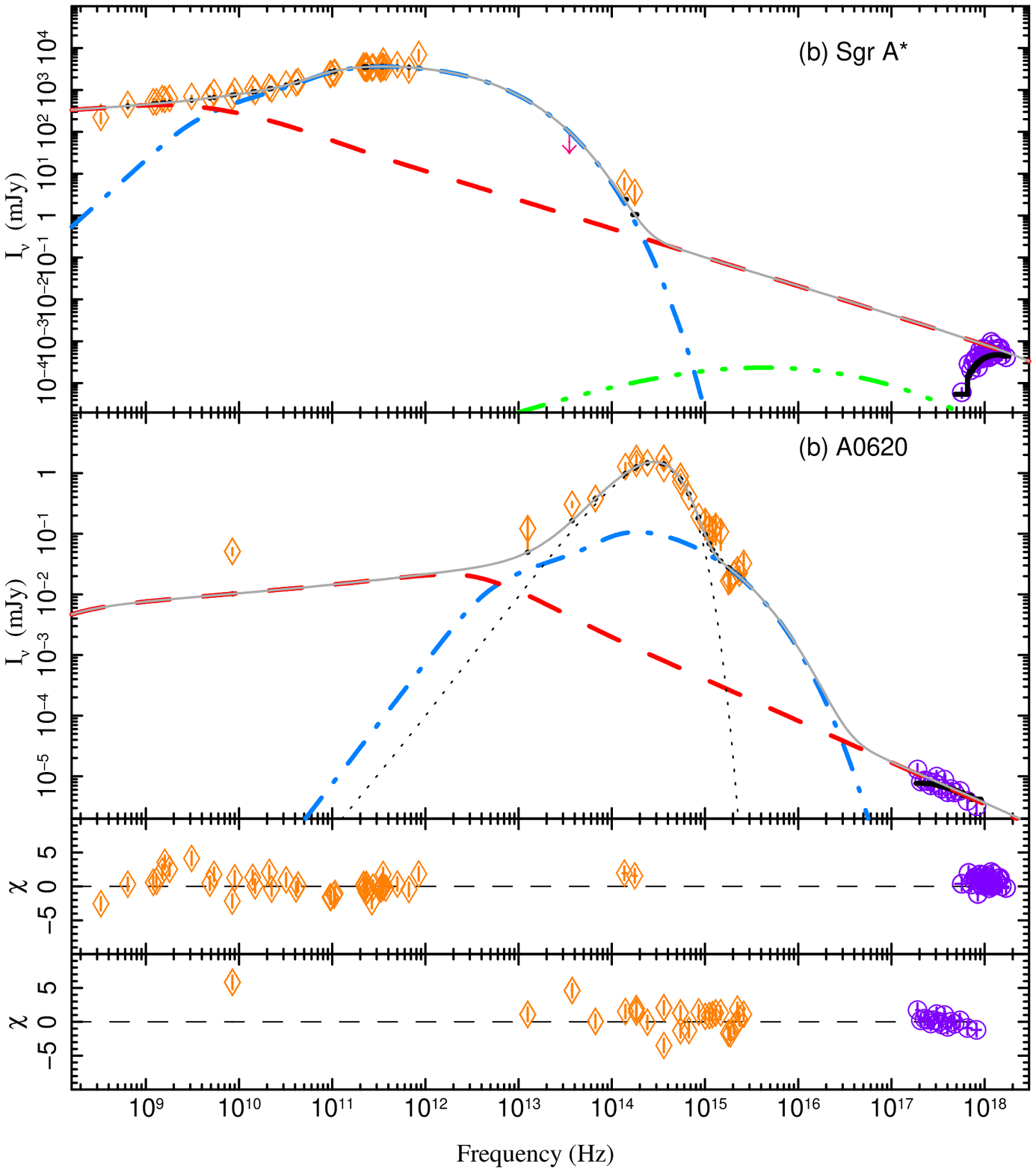}\hspace{0.1cm}\vspace{0.1cm}
\includegraphics[height=0.54\textwidth]{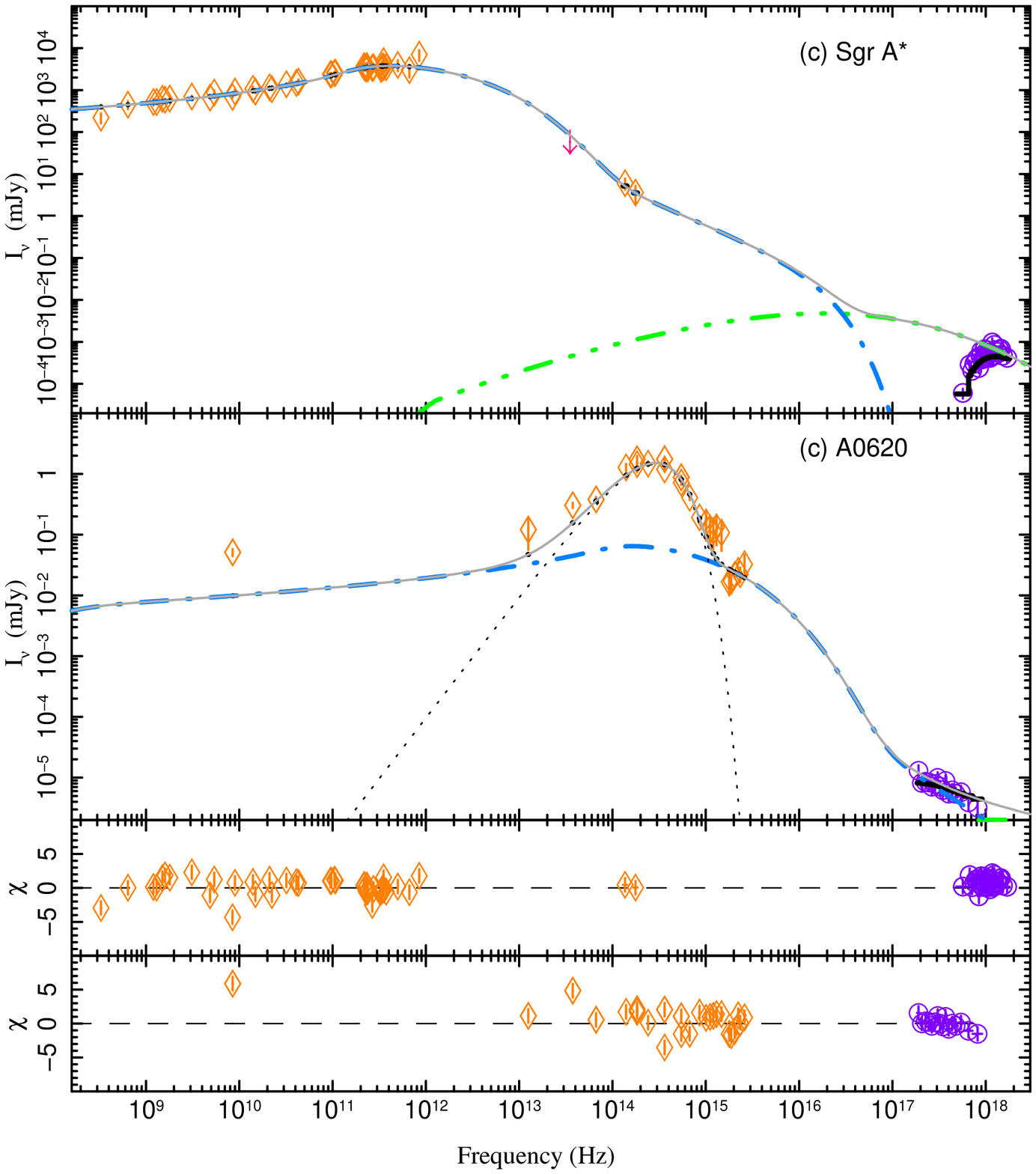}\hspace{0.4cm}\vspace{0.1cm}
\includegraphics[height=0.54\textwidth]{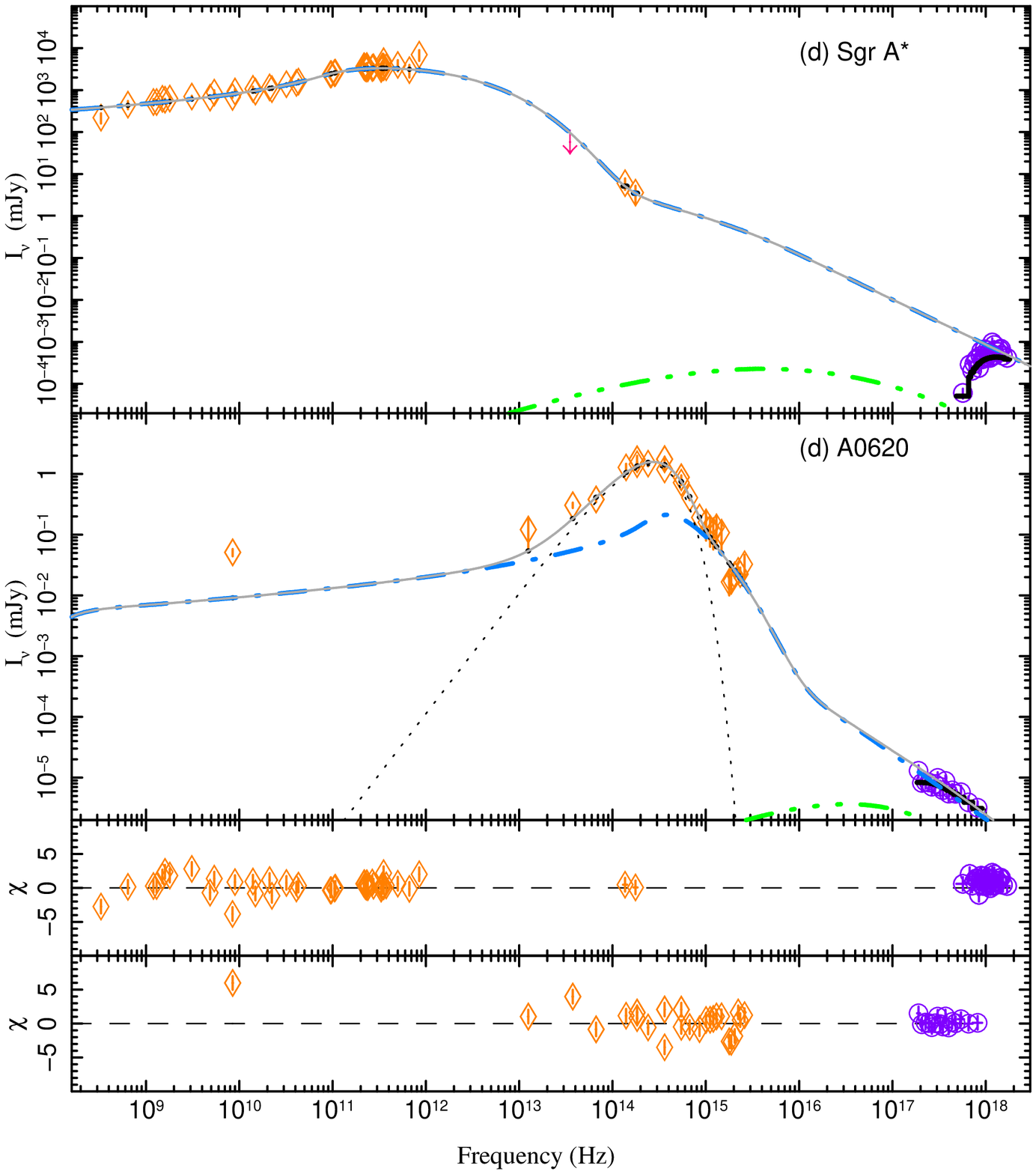}\hspace{0.01cm}\vspace{0.1cm}
  \ \caption{Joint spectral fits to both the~\sgra~and A0620-00 spectrum. The top left panel shows the case (a) model fit, top right the case (b) model fit, bottom left the case (c) model fit, and bottom right the case (d) model fit, where the~\sgra~and A0620 plots are indicated. In all A0620 plots the Orange diamonds show the radio - FUV spectrum loaded into ISIS as flux density measurements, and the 1st-order grating X-ray spectrum is shown with purple circles. In all~\sgra~plots the orange diamonds show radio - IR data loaded into ISIS as flux density measurements, with the X-ray spectrum in purple points. The model fit is indicated in black. In cases (a) and (b) the unabsorbed model components shown are pre-acceleration (thermal) synchrotron emission (blue dot-dashed line), post-acceleration synchrotron emission (red dashed line), SSC (green three-dot-dashed line), the blackbody spectrum of the stellar companion (black short-dashed line), and the total spectrum of \texttt{agnjet} (grey solid line). In cases (c) and (d) we do not include an explicit acceleration zone and thus there is no `post-acceleration` spectrum. Instead the synchrotron spectrum emitted by the full thermal + non-thermal electron distribution is shown with the blue dot-dashed line, and the SSC spectrum is shown by the green three-dot-dashed line. The bottom panels of each plot show the standardised residual photon counts. In cases (a) and (b) parameters $r_0$, $h_{ratio}$ (though this is fixed in case (b)), $z_{acc}$, $p$ and $k$ are tied, and in cases (c) and (d) we tie the same parameters except for $z_{acc}$, which does not function in these cases.}
  \label{fig:jointfits} % figure label for a0620 fits
 \end{figure*}
 
 \begin{figure*}
 \centering

 \includegraphics[height=0.21\textwidth]{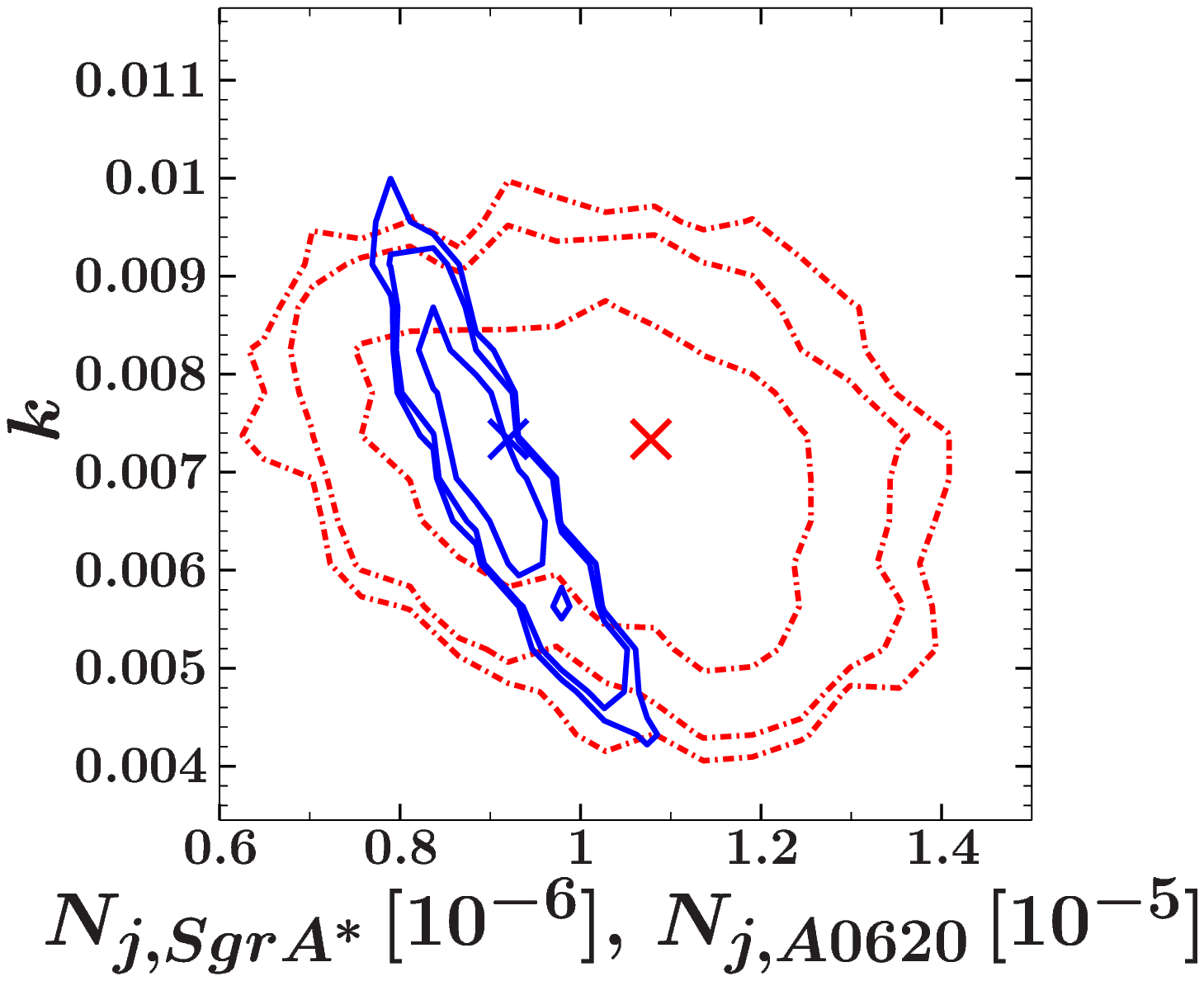}\hspace{0.1cm}\vspace{0.1cm}
 \includegraphics[height=0.21\textwidth]{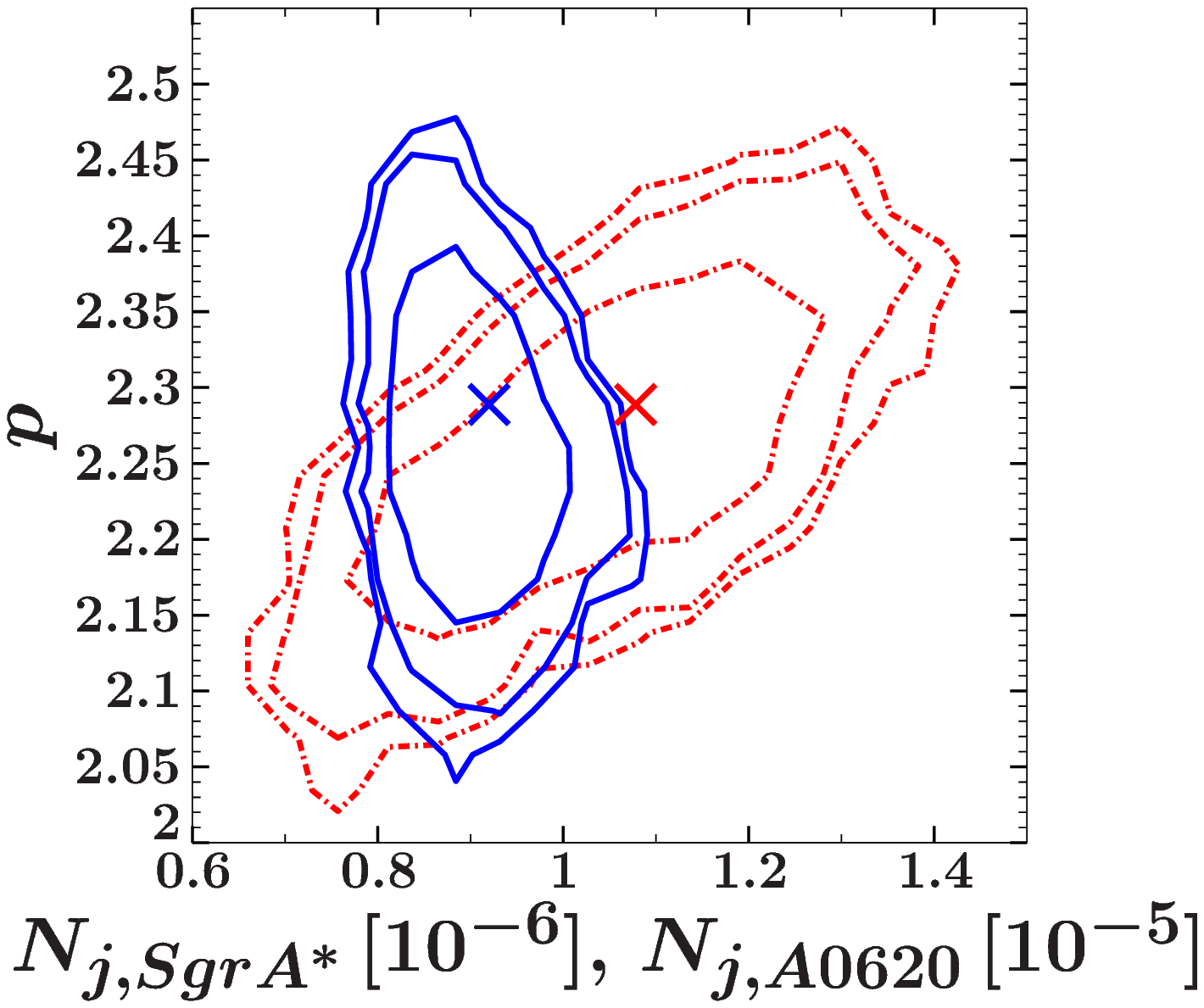}\hspace{0.1cm}\vspace{0.1cm}
 \includegraphics[height=0.21\textwidth]{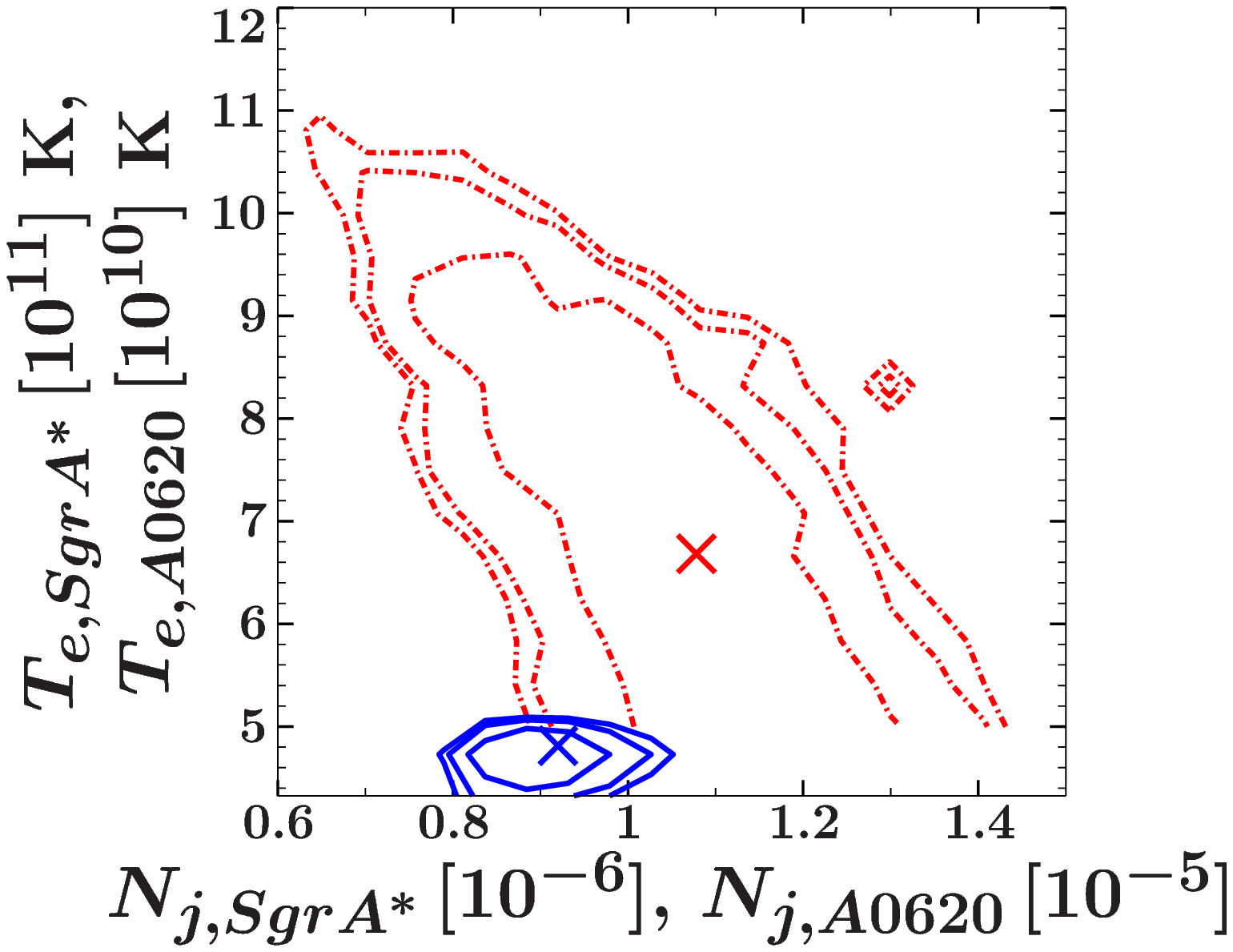}\hspace{0.1cm}\vspace{0.1cm} \\
 \includegraphics[height=0.21\textwidth]{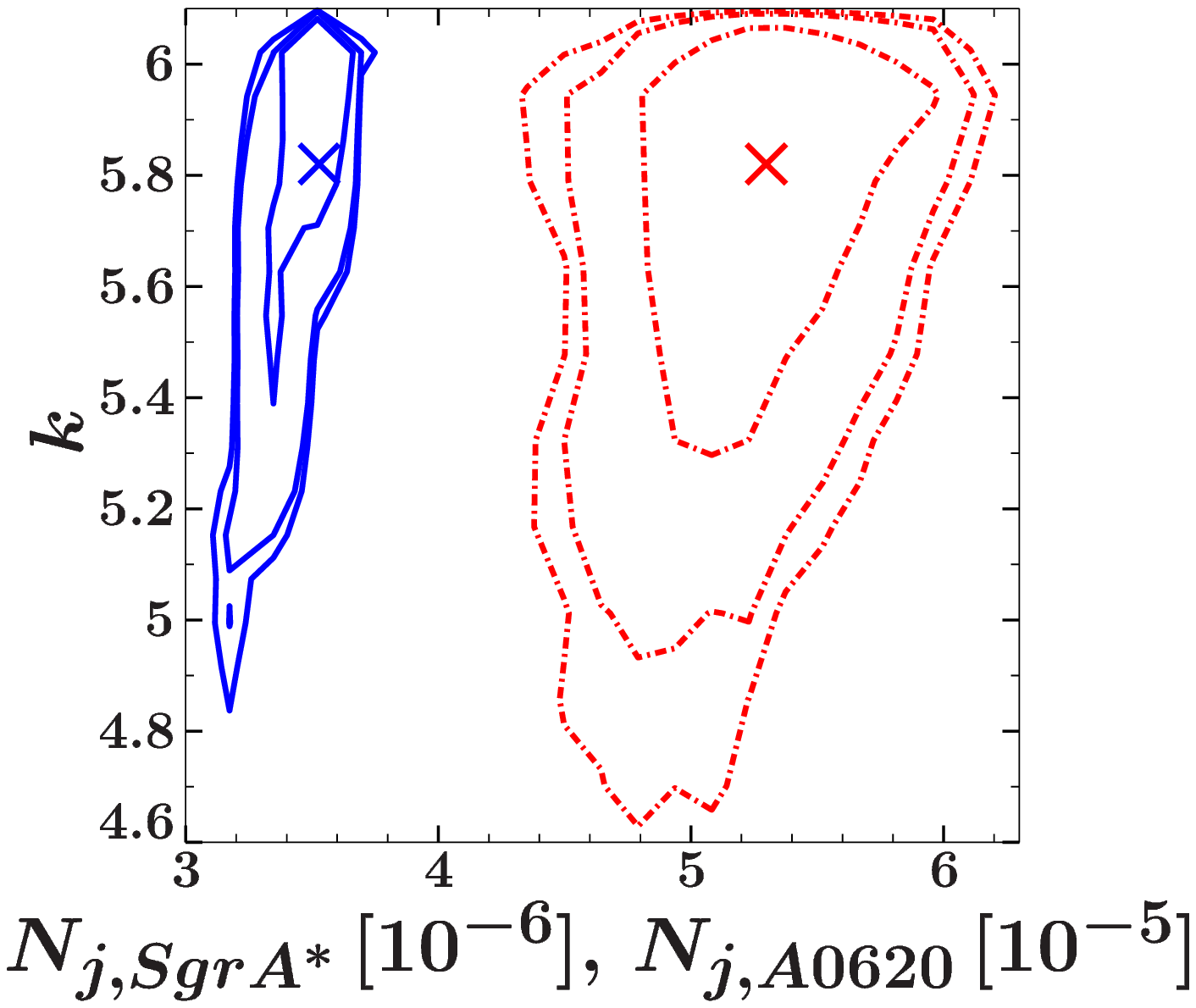}\hspace{0.1cm}\vspace{0.1cm}
 \includegraphics[height=0.21\textwidth]{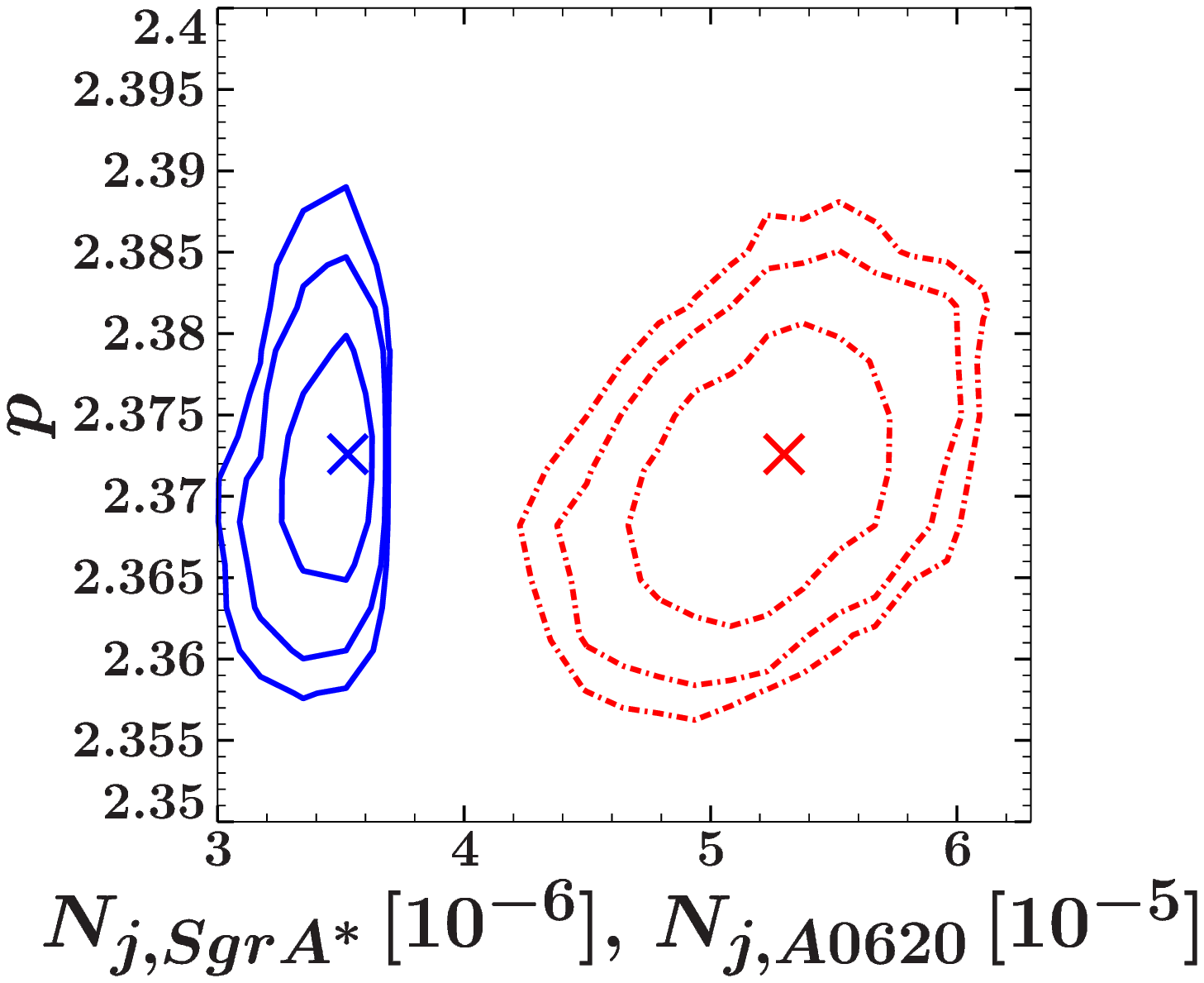}\hspace{0.1cm}\vspace{0.1cm}
 \includegraphics[height=0.21\textwidth]{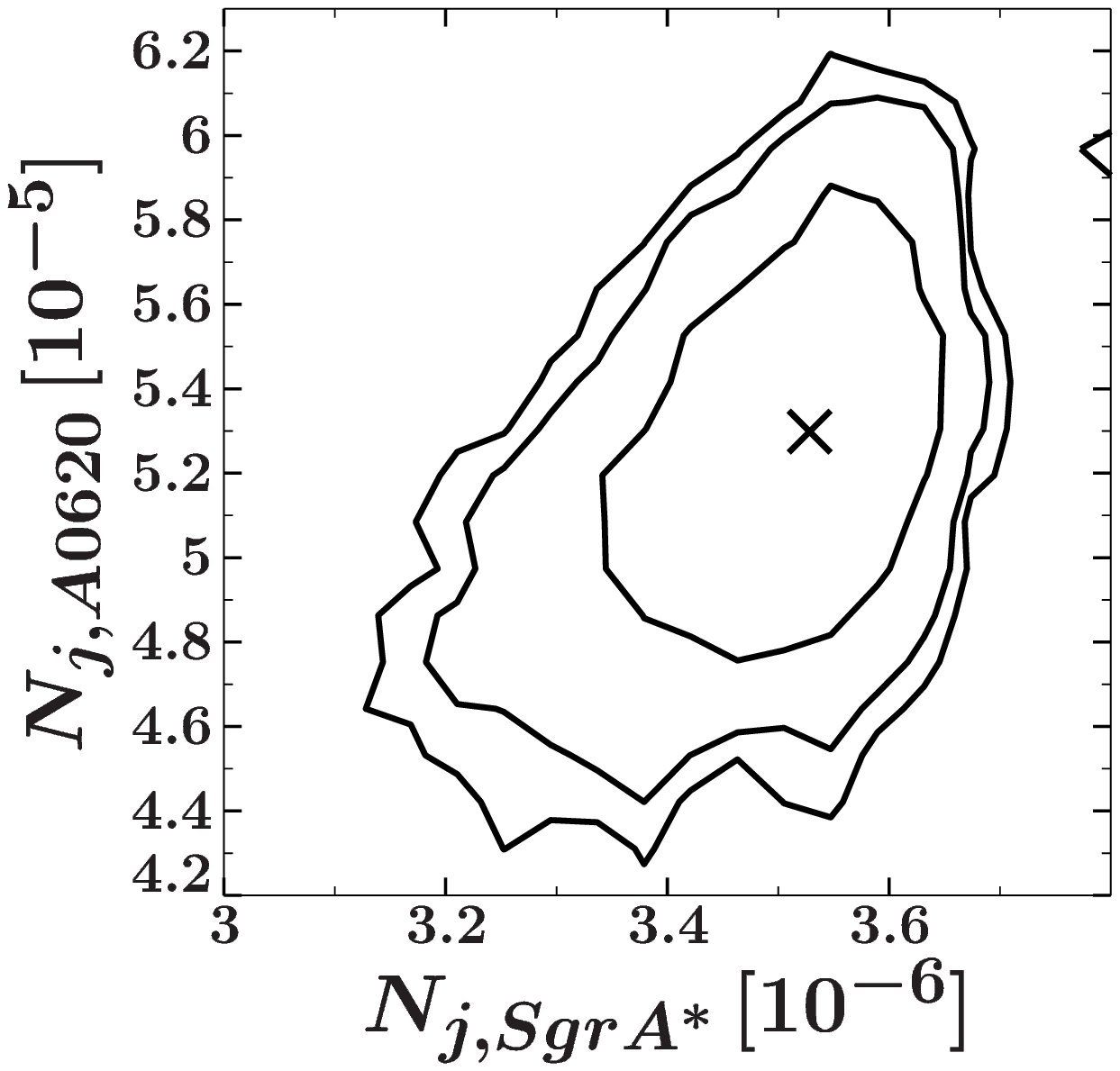}\hspace{0.1cm}\vspace{0.1cm} \\
 \includegraphics[height=0.21\textwidth]{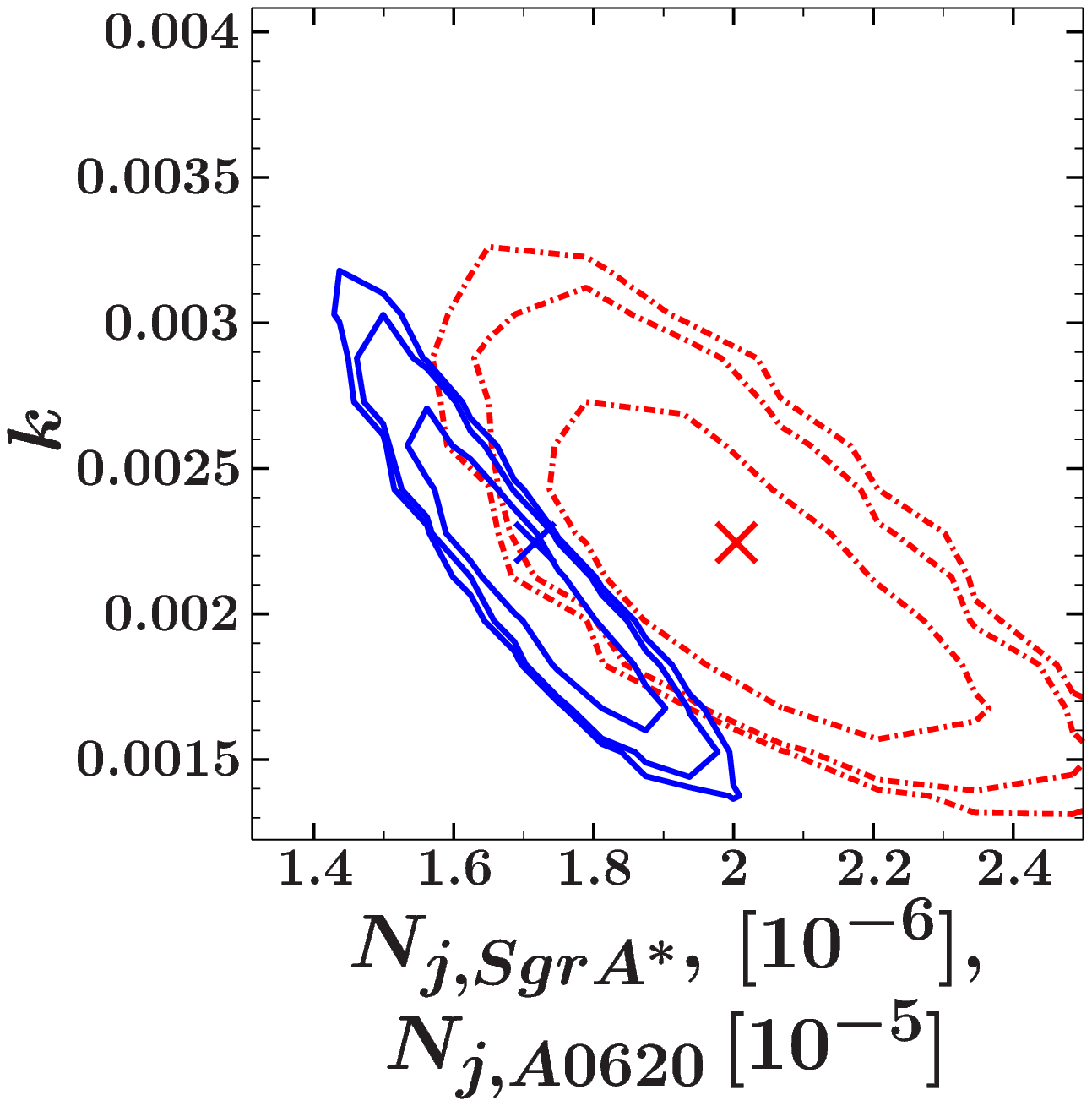}\hspace{0.1cm}\vspace{0.1cm}
  \includegraphics[height=0.21\textwidth]{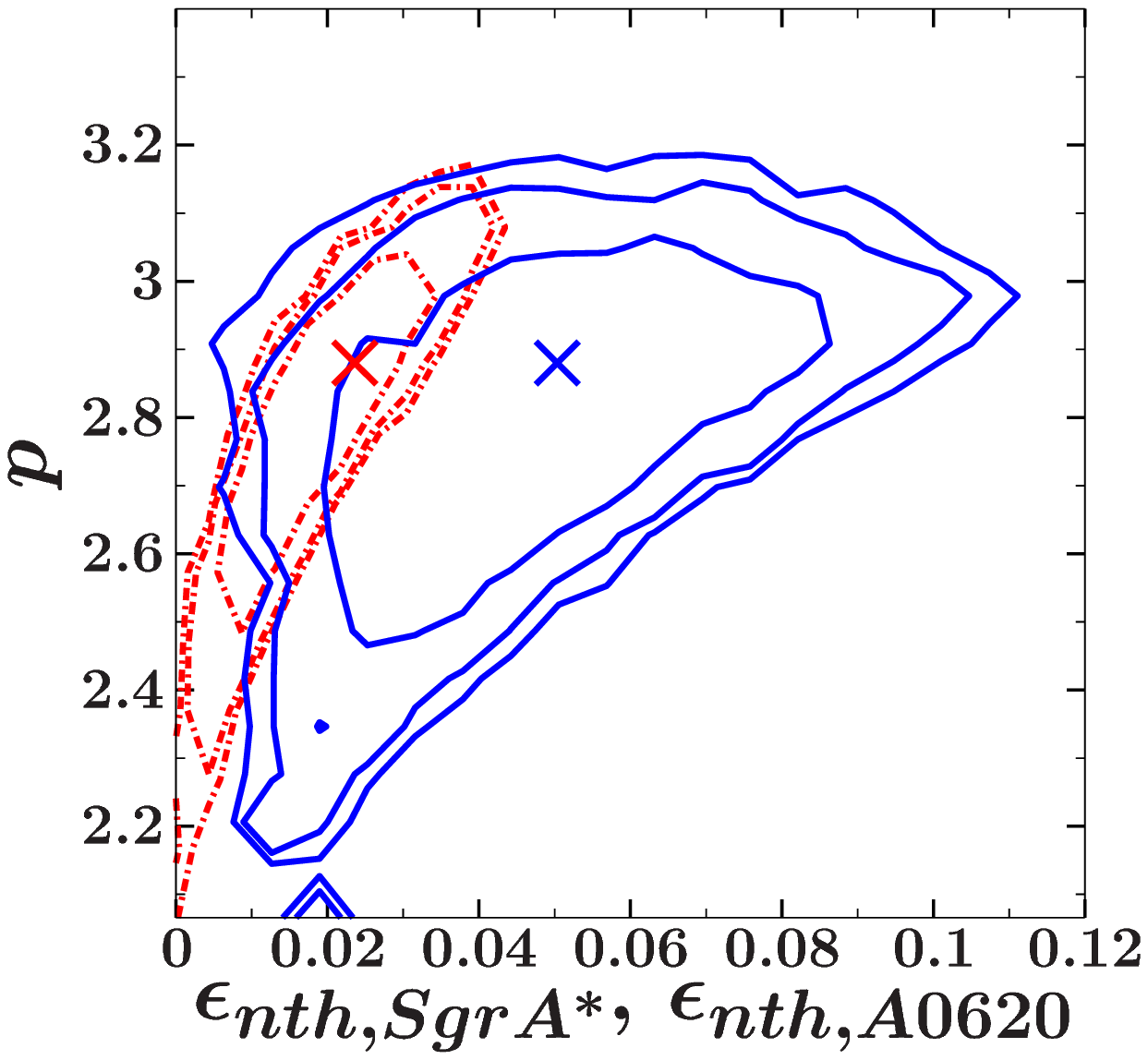}\hspace{0.1cm}\vspace{0.1cm}
 \includegraphics[height=0.21\textwidth]{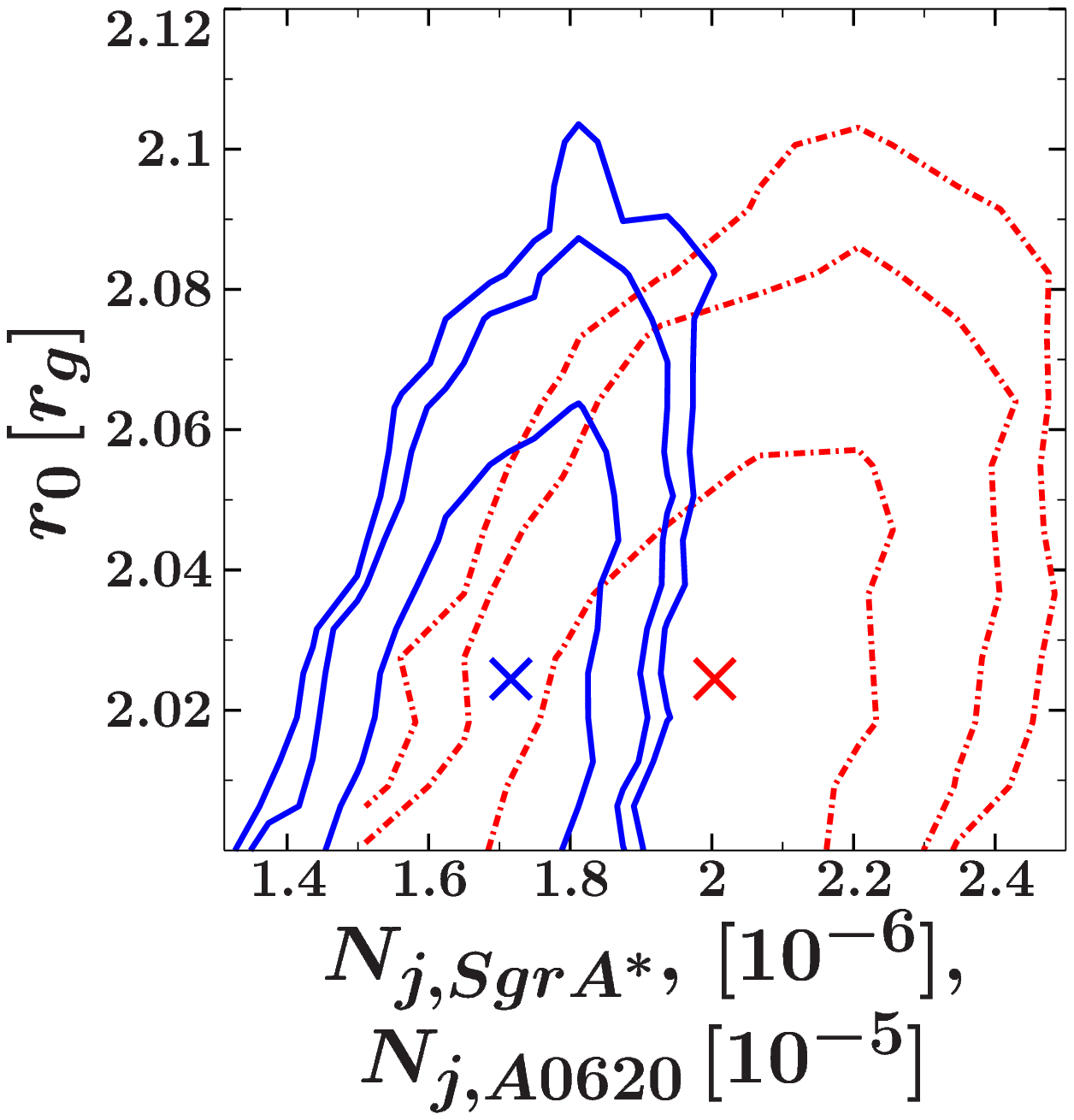}\hspace{0.1cm}\vspace{0.1cm}
 \includegraphics[height=0.21\textwidth]{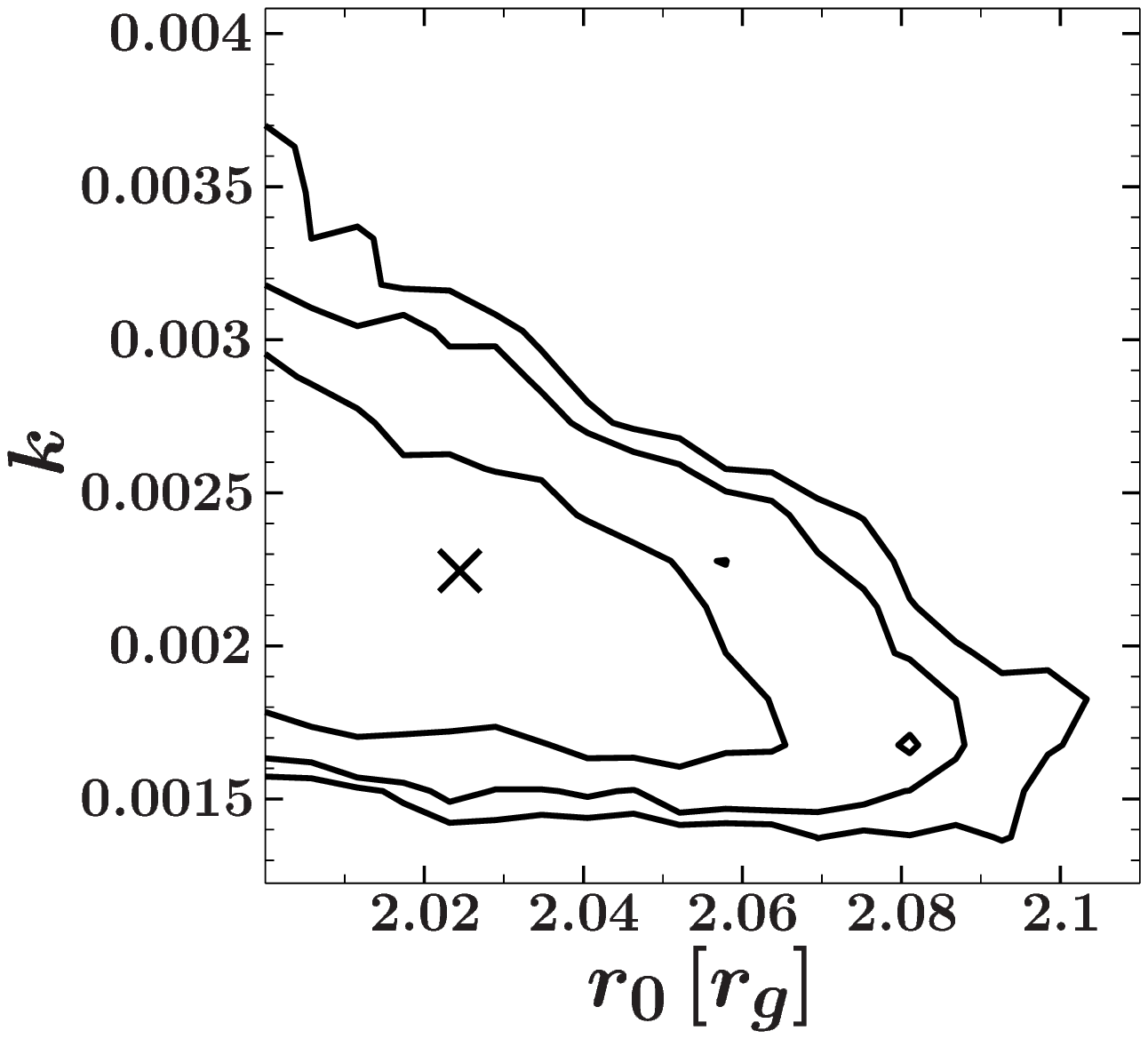}\hspace{0.1cm}\vspace{0.1cm}\\
  \includegraphics[height=0.21\textwidth]{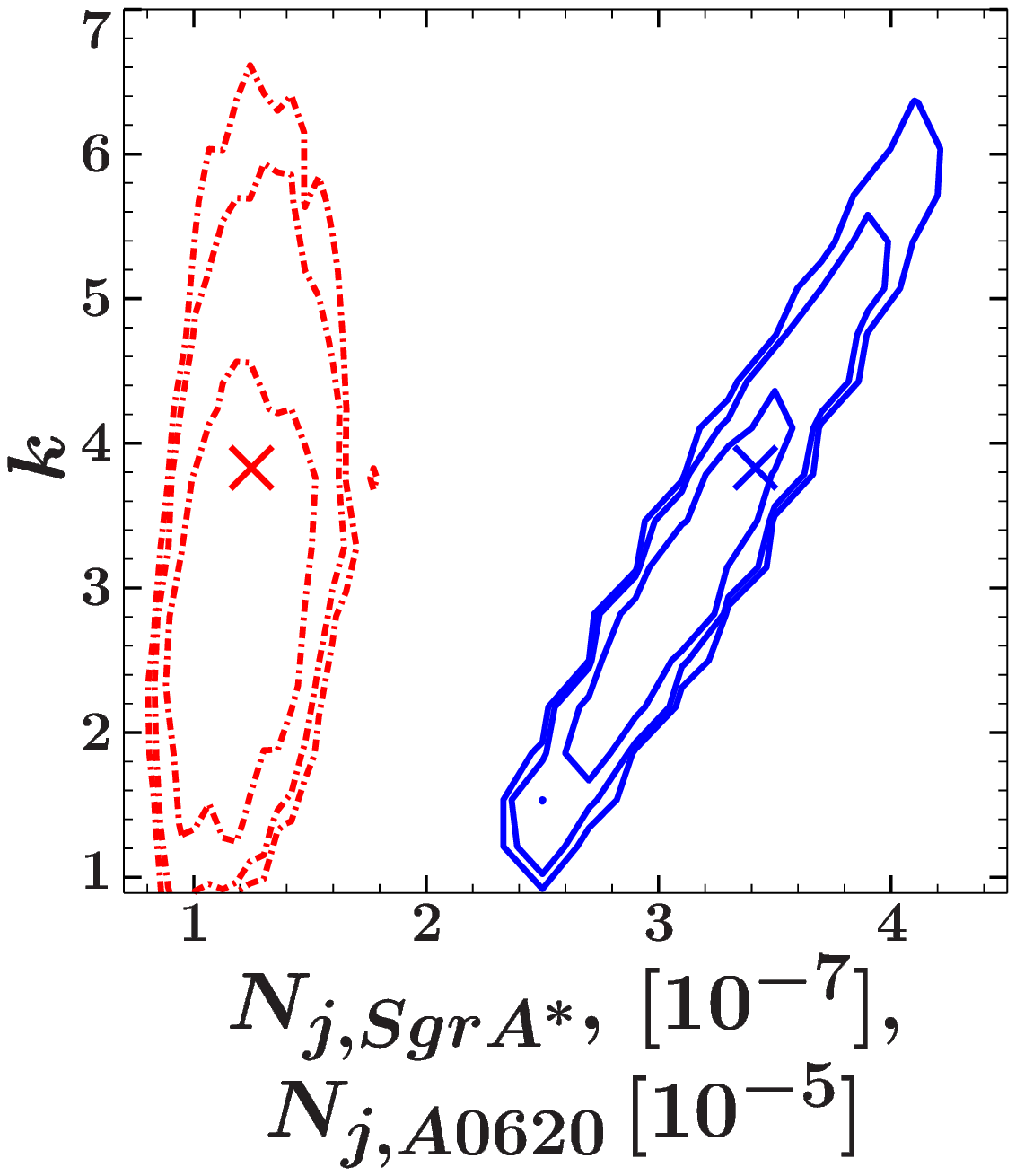}\hspace{0.1cm}\vspace{0.1cm} 
  \includegraphics[height=0.21\textwidth]{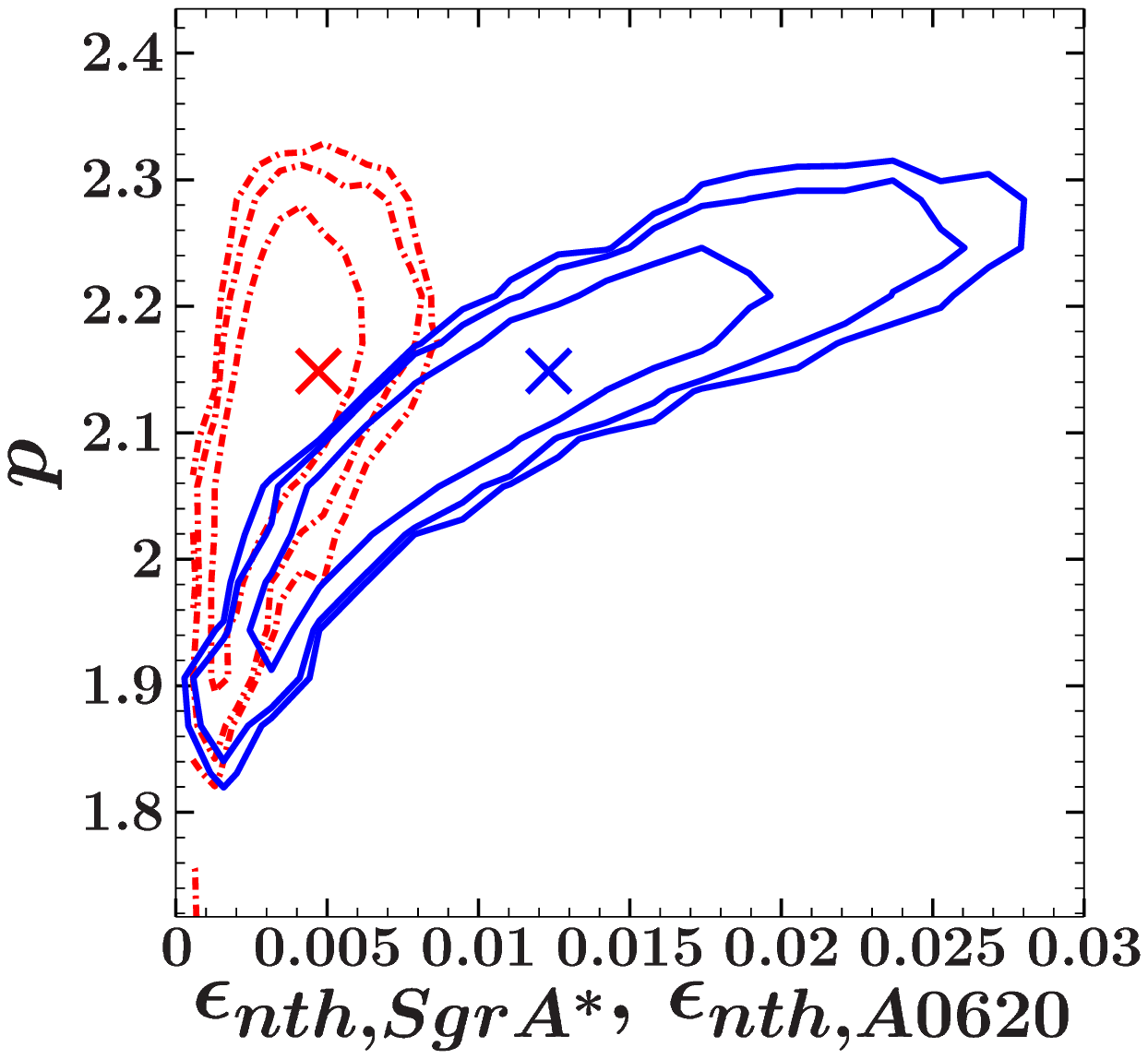}\hspace{0.1cm}\vspace{0.1cm} 
 \includegraphics[height=0.21\textwidth]{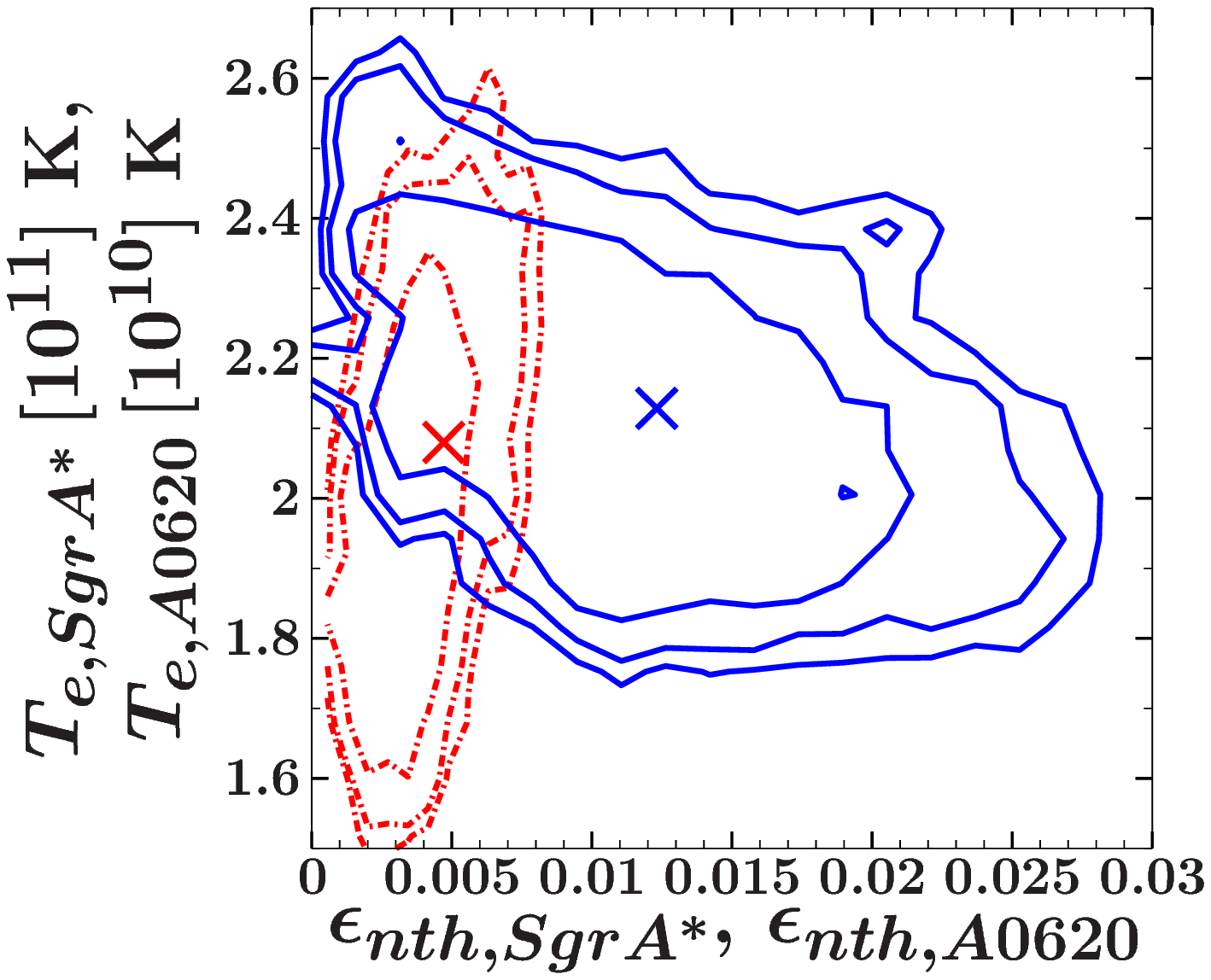}\hspace{0.1cm}\vspace{0.1cm}
 \includegraphics[height=0.21\textwidth]{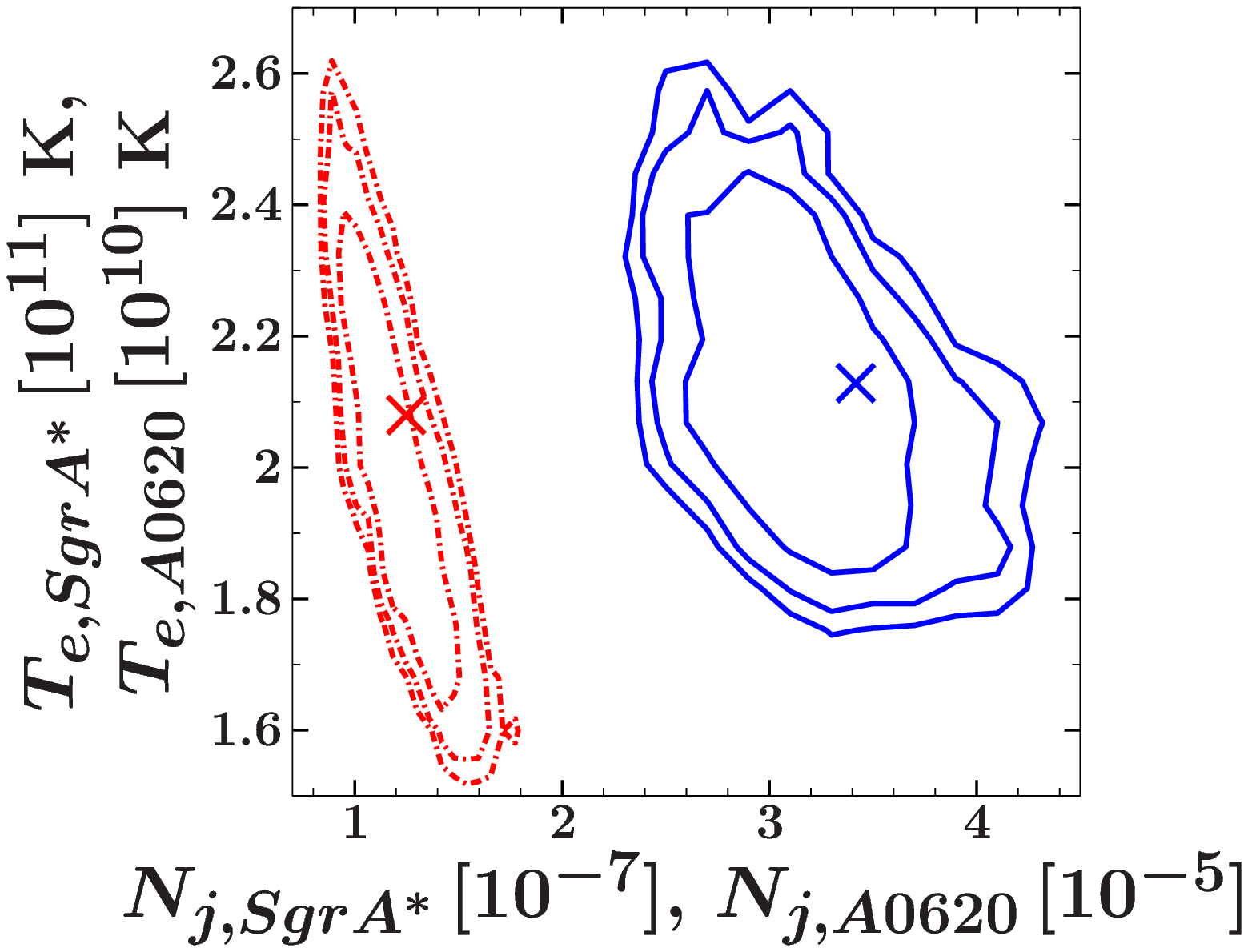}\hspace{0.1cm}\vspace{0.1cm}
 \caption{Two-dimensional contours of parameters of interest in joint fits to~\sgra~and A0620, covering all 4 cases, (a) - (d), shown consecutively from the top row of panels (case (a)) to the bottom row of panels (case (d)). Blue solid lines show contours at 0.68/0.90/0.95 confidence for~\sgra~fitted parameters, and red dotted lines show the equivalent for A0620 parameters. Black solid lines show contours either for joint-fitted parameters or comparisons between a parameter applying to each source individually. The crosses indicate the MLEs.}
 \label{fig:conf}
 \end{figure*}

  % TABLE - A0620-00 &~\sgra~joint-fitting params, thermal dist %%%%%%%%%%%%%%%%%%%%%%%%%%%%%%%%%%%%%%%%%%%%%%%%%%%%%%%%%%%
 \begin{table*}
 \centering
  \caption{Fitted parameters for Synchrotron-and-SSC-dominated joint spectral fits to A0620 and \sgra. Shown are 4 model cases for fits, (a) thermal particle injection, SSC-dominated, (b) thermal particle injection, synchrotron-dominated, (c) mixed particle injection, SSC-dominated, (d) mixed particle injection, synchrotron-dominated. Confidence limits are at the 90\% level, a result of our MCMC exploration of the posterior distributions of the parameters. The resultant MLEs are given by the median point of each posterior distribution - this proves to be a good measure of the converged best-fit of the MCMC routine. The final column shows the resultant $\chi^2$ and the degrees of freedom (DoF). From left-to-right the following parameters are shown: $N_H$, the Hydrogen column density along the line-of-sight to the source, $N_j$, the jet power, $p$, the spectral index of the power-law-distributed electrons, $T_e$, the temperature of the electron distribution (Maxwell-J\"{u}ttner), $z_{acc}$, the location of acceleration in the jet (only applicable when a pure thermal particle distribution is injected at the base, cases (a) and (b)), $r_0$, the radius of the jet nozzle, $h_{ratio}$ the ratio of the nozzle height $h_0$ to the jet-base radius $r_0$, $k$, the energy partition factor, and $\epsilon_{nth}$, the fraction of energy density in non-thermal electrons. The parameters $r_0$, $h_{ratio}$, $z_{acc}$, $p$, and $k$ are tied together where applicable (in cases (c) and (d) there is no acceleration zone in the jet, and thus $z_{acc}$ is null.) Jet-base electron densities ($n_{e,0}$) and magnetic field strengths ($B_0$) are also shown for the corresponding MLEs.}
 \begin{tabular*}{1.02\textwidth}{p{0.8cm} p{1cm} p{1cm} p{1cm} p{0.8cm} p{0.8cm} p{1.1cm} p{1cm} p{1.3cm} p{0.7cm} p{0.8cm} p{1.2cm} p{1.2cm}}
  \hline
 \hline
 \textbf{Source}  &  \boldmath{$N_H$}  & \boldmath{$N_j$}  & \boldmath{$p$} & \boldmath{$T_e$}  & \boldmath{$z_{acc}$} & \boldmath{$r_0$}  & \boldmath{$h_{ratio}$}  & \boldmath{$k$} & \boldmath{$\epsilon_{nth}$} & \boldmath{$\frac{\chi^2}{DoF}$} &  \boldmath{$n_{e,0}$} & \boldmath{$B_0$}\\
  & [$10^{22}$ cm$^{-2}$] & [$10^{-7}$] & & [$10^{10}$ K] & [$r_g$]  & [$r_g$] &  [$h_0/r_0$]&  & [$10^{-2}$]& & [cm$^{-3}$] & [G] \\
 \\
 \hline
& & & & & & \textbf{(a)} \\
 \hline
 \\
 ~\sgra~ & $11.8^{+0.6}_{-0.7}$ & $9.2^{+1.2}_{-0.8}$ & ... & $48^{+2}_{-2}$ & ... & ...  & ... & ... & ... & ... & $\mathit{8.3\times10^{6}}$ & $\mathit{17.4}$\\
 \\
 A0620  & $0.15^{+0.19}_{-0.05}$ & $1100^{+200}_{-300}$ & ... & $0.7^{+0.3}_{-0.2}$  & ... & ... & ... & ... & ... & ... & $\mathit{3.8\times10^{16}}$ & $\mathit{1.5\times10^5}$\\
 \\
 Joint & ... & ... & $2.3^{+0.1}_{-0.2}$ & ... & $20^{+9}_{-5}$ & $2.27^{+0.08}_{-0.19}$ & $1.22^{+0.12}_{-0.07}$ & $0.007^{+0.002}_{-0.002}$ &  ... & 329/219 & ... & ...\\
 \\
  \hline
 & & & & & &  \textbf{(b)} \\
  \hline
  \\
~\sgra~& $12.2^{+0.7}_{-0.7}$ & $3.5^{+0.1}_{-0.2}$ & ... & $20.9^{+0.1}_{-0.5}$ &  ... & ... & ...  & ... & ... & ... & $\mathit{3.5\times10^5}$ & $\mathit{66.9}$\\
 \\
 A0620 & $0.14^{+0.09}_{-0.03}$ & $53^{+6}_{-6}$ & ... & $4.7^{+1.4}_{-0.7}$ & ... & ... & ... & ... & ...  & ... & $\mathit{1.4\times10^{13}}$ & $\mathit{2.0\times10^5}$\\
 \\
 Joint & ... & ... & $2.37^{+0.01}_{-0.01}$ & ... & $180^{+10}_{-20}$ & $4.0^{+0.3}_{-0.2}$ & $1.5^f$ & $5.8^{+0.3}_{-0.8}$ & ... & 406/220 & ... & ...\\
 \\
 \hline
 & & & & & & \textbf{(c)} \\
 \hline
 \\
~\sgra~& $12.3^{+0.7}_{-0.7}$ & $17^{+2}_{-2}$ & ... & $41.7^{+0.3}_{-0.8}$ &  ... & ... & ...  & ... & $5^{+4}_{-3}$ & ... & $\mathit{2.3\times10^7}$ & $\mathit{14.8}$\\
 \\
 A0620 & $0.14^{+0.09}_{-0.03}$ & $200^{+40}_{-20}$ & ... & $17^{+2}_{-5}$ & ... & ... & ... & ... & $<3$  & ... & $\mathit{3.8\times10^{14}}$ & $\mathit{3.9\times10^{4}}$\\
 \\
 Joint & ... & ... & $2.9^{+0.2}_{-0.4}$ & ... & ... & $2.02^{+0.06}_{-0.02}$ & $0.85^{+0.05}_{-0.03}$ & $0.002^{+0.001}_{-0.001}$ & ... & 390/218 & ... & ...\\
 \\
 \hline
& & & & & &  \textbf{(d)} \\
 \hline
 \\
~\sgra~& $13.4^{+0.9}_{-0.8}$ & $3.4^{+0.7}_{-0.8}$ & ... & $21^{+3}_{-3}$ &  ... & ... & ...  & ...& $1.2^{+1.2}_{-0.9}$ & ... & $\mathit{5.4\times10^5}$ & $\mathit{67.7}$\\
 \\
 A0620 & $0.20^{+0.11}_{-0.09}$ & $120^{+30}_{-30}$ & ... & $2.1^{+0.4}_{-0.4}$ & ... & ... & ... & ... & $0.5^{+0.3}_{-0.3}$ & ... & $\mathit{1.2\times10^{14}}$ & $\mathit{3.2\times10^5}$ \\
 \\
 Joint & ... & ... & $2.1^{+0.1}_{-0.2}$ & ... & ... & $3.7^{+0.4}_{-0.3}$ & $1.08^{+0.12}_{-0.07}$ & $4^{+2}_{-2}$ & ...& 381/218 & ... & ...\\
 \\
 \hline
 \end{tabular*}
 \label{tab:jointfits}
 \textbf{Notes.} $^f$ Frozen parameter
 \end{table*} 

 %%%%%%%%%
 % SUBSECTION - joint fits - case a %%%%%%%%%%%%%%%%%%%%%%%%%%%%%%%%%%%%%%%%%%%%%%%%%%%%%
 %%%%%%%%%
 \subsection{Case (a): thermal particle injection, SSC-dominated}
 \label{subsec:joint_case_a}
Here we achieve good fits to the data given a very sub-equipartition magnetic field, but the electron temperature still needs to be high in order to produce the required X-ray flux via pure SSC emission. The fit to A0620 sees a statistically significant decrease in the jet-base radius compared with individual spectral fits, whilst this value is consistent with single fits of this case to \sgra (see section \ref{sec:singlefits}). This property applies independently of the type of model we are fitting (though SSC-dominated fits, i.e. cases (a) and (c), give slightly more compact jet bases). The electron temperature of A0620's inner accretion flow also decreases, and the particle acceleration zone ($z_{acc}$) drops compared with individual fits. A significant change is seen for the jet power of A0620, which has to increase to account for the drop in the energy partition parameter (and thus a reduced magnetic field strength), an effect displayed clearly in Figure \ref{fig:conf}, showing the two-dimensional confidence contours of parameters in which we see correlations. 
 
 %%%%%%%%%
 % SUBSECTION - joint fits - case b %%%%%%%%%%%%%%%%%%%%%%%%%%%%%%%%%%%%%%%%%%%%%%%%%%%%%
 %%%%%%%%% 
 \subsection{Case (b): thermal particle injection, synchrotron-dominated}
 \label{subsec:joint_case_b}
 Here we see that the result of tying $z_{acc}$ during synchrotron-dominated fits is to push its value higher to allow the A0620 X-ray spectrum to be modelled sufficiently, and this also results in tighter constraints on its value. Solutions in which particle acceleration occurs so distant from the black hole are unlikely to be the source of bright X-ray flares given the timescale constraints on \sgra's X-ray variability \citep{nei13,bar14}; we expect the flare emission to be originating from within $\sim5~r_g$ of the black hole during the brightest flares. One notices a few correlations in the 2D contours shown in Figure \ref{fig:conf}, including a weak but persistent positive correlation between $N_j$ and $k$ for~\sgra~and A0620 during these synchrotron-dominated states. This implies that by providing energy to the magnetic field in \sgra, the drop in electron density is enough to force an increase in the injected power; fits to \sgra, even in the synchrotron-dominated case, are very sensitive to electron density. 

 %%%%%%%%%
 % SUBSECTION - joint fits - case c %%%%%%%%%%%%%%%%%%%%%%%%%%%%%%%%%%%%%%%%%%%%%%%%%%%%%
 %%%%%%%%%
 \subsection{Case (c): mixed particle injection, SSC-dominated}
 \label{subsec:joint_case_c}
 Figure \ref{fig:jointfits} shows, similar to case (a), that the mid-IR constraints on the jet-base electron temperature result in a significantly reduced partition factor. This is consistent with the case (c) single fit to \sgra, but significantly lower than single fits to A0620. The electron temperature of~\sgra~is high, even with such a sub-equipartition flow, and the mid-IR limits are almost surpassed. The electron temperature of~\sgra~remains consistent with single fits, but we see a statistically significant increase in the electron temperature of A0620, again reflecting the evolution to a highly sub-equipartition magnetic field (reflected in the jet-base magnetic field values shown in Table \ref{tab:jointfits}). We also note that whilst we see a correlation between $r_0$ and $T_e$ in single fits to A0620, this correlation is not present in our joint fits, a possible indication that the joint fitting approach indeed reduces some of the physical degeneracy. The value of $\epsilon_{nth}$ is statistically indistinguishable between~\sgra~and A0620, though we note that for A0620 its value is constrained to lower fractions than those in the fit to~\sgra. There is a $k$--$r_0$ weak anti-correlation introduced by performing this joint fit, which may reflect the relative importance of the energy partition over the electron temperature (this is evident also from the weak constraints we are able to place on $T_e$). We also note that, as shown in Figure \ref{fig:jointfits}, the X-ray spectrum of A0620 is dominated by the synchrotron emission from the non-thermal tail injected at the base of the jet, indicating that in this case the mass-scaling changes the phenomenology of this class of fit to A0620.
 
 %%%%%%%%%
 % SUBSECTION - joint fits - case d %%%%%%%%%%%%%%%%%%%%%%%%%%%%%%%%%%%%%%%%%%%%%%%%%%%%%
 %%%%%%%%%
 \subsection{Case (d): mixed particle injection, synchrotron-dominated}
  \label{subsec:joint_case_d}
Here we see a fast-cooling dominated synchrotron spectrum, produced by a population of accelerated electrons close to the black hole, consistent with the observed timescales of the flares \citep{bar14}. \cite{rus13} indicate that BHBs that decline into quiescence seem to exhibit a cooling-break evolution down to UV energies. We find cooling breaks for both~\sgra~and A0620 at $\sim10^{16}$ Hz and $\sim10^{17}$ Hz respectively. Whilst the values of $p$ and $\epsilon_{nth}$ remain consistent with single fit values for both~\sgra~and A0620, other parameters show statistically significant changes. \sgra's jet power is decreased, $T_e$ and $r_0$ drop in the A0620 fit, and \sgra's inner flow goes to slightly lower energy partition, decreasing the jet-base magnetic field. We note that just as in case (c), the $T_e$ - $r_0$ correlation in A0620 fits is removed when fitting jointly, however Figure \ref{fig:conf} shows that a correlation is introduced between $T_e$ and $N_j$ for both~\sgra~and A0620.

 %%%% DISCUSSION/CONCLUSION %%%%%%%%%%
 \section{Discussion and Conclusion}
 \label{sec:conclusion}
We have shown that mass-scaling in weakly accreting black holes (the FP) can be exploited to break degeneracies in spectral modelling. This has been explored for higher luminosity, hard-state-like sources by \cite{mar15}; here we extend this study for the first time to the most quiescent sources. We find that the majority of key parameters can be tied in the modelling of~\sgra~and A0620, namely the jet-base radius $r_0$, the distance from the black hole along the jet axis at which particle acceleration occurs $z_{acc}$, the height of the jet nozzle $h_0/r_0$, the partition of energy between the magnetic field and electrons $k$, and the spectral index of the accelerated electron distribution $p$. This new approach reduces the strength of some parameter correlations (i.e. degeneracy) and better constrains those tied parameters, leading to more informative distinctions between different model classes. \\
\indent Our results imply that if the X-ray spectrum of~\sgra~during the brightest flares is dominated by SSC emission from a purely thermal population of electrons, its jet-base magnetic field must be significantly sub-equipartition, and the electron temperatures push to high values, giving rise to mid-IR fluxes close to the upper limit obtained by \cite{hau12}. If indeed A0620 and~\sgra~can be related via their energy partition, then we would expect sources accreting at $l_{X}\sim10^{-9}$--$10^{-8}$ to be highly sub-equipartition. \\
\indent An alternative physical scenario in which the X-ray spectra of A0620 and~\sgra~are dominated by synchrotron emission from an injected non-thermal distribution of electrons is consistent with what we know about the variability of~\sgra~(as opposed to the solutions we find in which particle acceleration occurs in regions $>5~r_g$ from the black hole \citealt{bar14}), and can also explain the broadband spectra of both sources. In this scenario, a small fraction ($\sim$ a few \%, statistically consistent for both~\sgra~and A0620) of the injected electron energy density is in non-thermal electrons, and these electrons experience a cooling break at $\sim10^{16}$ Hz and $\sim10^{17}$ Hz for~\sgra~and A0620 respectively. These results are consistent with previous modelling of~\sgra~that indicates a cooling break in the synchrotron spectrum between IR and X-ray energies \citep{dib12,dib14}. We therefore find that the suggestion of \cite{mark10} and \cite{plot15} that quiescent BHBs enter a regime of inefficient particle acceleration is an inherently degenerate one. The effect of inefficient acceleration (and therefore SSC-dominated X-ray states) can be subsumed by efficient particle cooling within the peak synchrotron emission zones. This interpretation is consistent with the findings of \cite{rus13} that as BHBs evolve to quiescence their cooling breaks evolve to lower frequencies such that the break may be observed in the UV band. We also note that recent single-zone modelling of the inner accretion flow of~\sgra, with the goal of reproducing the observed IR/X-ray flare distributions, shows that particle acceleration as well as density and magnetic field changes are key to producing the flares \citep{dib16}. \\
\indent Both these scenarios are consistent with a matter-dominated disc-jet system, in which the jet magnetisation is low \citep{bp82}, as opposed to a Poynting-flux dominated jet in which we should expect high magnetisations \citep{bz77}. As discussed in section \ref{sec:model}, we are unable to properly describe a highly magnetised jet with our model, since we explicitly assume $U_B\le{nm_pc^2}$. However, we note that a scenario in which the jet has a Poynting-dominated spine surrounded by a matter-dominated outer sheath may still be consistent with the conditions we find \citep{hk06,mos16}.\\
\indent Achieving efficient particle acceleration, whether via internal shocks or magnetic reconnection, is an ongoing area of study, with some recent progress coming from PIC simulations of both such mechanisms (e.g. \citealt{ss11,ss14,ssa13,spg15}). In the shock-acceleration scenario (assuming quasi-perpendicular shocks), it proves difficult to accelerate electrons to high energies unless the pre-shock conditions are at low magnetisation. Conversely in the magnetic reconnection scenario, electrons may be accelerated to high energies if the inner regions are highly magnetised. We propose that at the most quiescent levels ($l_{X}\sim10^{-9}$--$10^{-8}$) accreting black holes struggle to achieve the structures necessary for efficient particle acceleration (as represented by the acceleration regions at $z>z_{acc}$), but that there are still likely a small fraction of non-thermal radiating particles at the jet base (in the fast-cooling regime) of the black hole produced via another acceleration mechanism \citep{yqn03}. We also propose that such sub-equipartition conditions at the jet base favour brighter SSC emission in the X-rays, which reinforces the findings of \cite{plot15} that a switch to quiescence in BHBs is associated with a compact jet base (a few gravitational radii) and sub-equipartition magnetic fields with respect to the electrons. These sub-equipartition magnetic fields may coincide with the production of weakly magnetised outflows as opposed to highly collimated jets. \\
 \indent We can understand more about what each mechanism (synchrotron or SSC) predicts in terms of the emission timescales of \sgra's daily flares, in particular the connection between X-ray and IR emission, by considering the timescales for particle acceleration in weakly relativistic outflows. An electron with a Lorentz factor $\gamma$ gyrating in a magnetic field $B$, will have a peak synchrotron frequency of
\begin{equation}
\label{eq:synch_freq}\nu = \frac{q\gamma^2B}{2\mathrm{\pi}m_e c}
\quad \Rightarrow \quad
\gamma \approx 3\times10^4 \left(\frac{B}{100\ \rm G}\right)^{-1/2}\left(\frac{h\nu}{\rm 1 \ keV}\right)^{1/2}.
\end{equation}
From Equation \ref{eq:synch_freq}, we can see that electrons with energies $\sim 15$ GeV will be capable of radiating X-rays in a 100 G strength magnetic field.

The time it takes for an electron to be accelerated to an energy capable of radiating at a given frequency $\nu$ in diffusive shock acceleration (DSA) is (e.g., \citealt{cs14})
\begin{equation}
\label{eq:t_time}
t_{\rm acc}= \frac{6 D(\epsilon)}{v_s^2},
\end{equation}
where the $v_s$ is the velocity of the upstream flow in the downstream reference frame relative to the shock. If we assume that the particle acceleration happens in the Bohm diffusion limit, the diffusion coefficient is (e.g., \citealt{jok87,am15})
\begin{equation}
\label{eq:Bohm}
D_B(\epsilon)=  \frac{c r_L}{3}; \quad r_L = \frac{\gamma m c^2}{qB}.
\end{equation}

Combining Equations (\ref{eq:synch_freq}), (\ref{eq:t_time}), and (\ref{eq:Bohm}) yields the following acceleration time:

\be
\label{eq:t_acc}
\begin{split}
t_{\rm acc} &= \frac{2c^2}{v_s^2}\frac{r_L}{c} \\
& \approx 1.3\times 10^{-4} \left(\frac{B}{100\ \rm G}\right)^{-3/2}\left(\frac{h \nu}{1\ \rm keV}\right)^{1/2}\left(\frac{v_s}{0.5c}\right)^{-2} \ \mathrm{s}.
\end{split}
\ee

We assumed shock acceleration to derive Equation (\ref{eq:t_acc}), but a derivation assuming magnetic reconnection would yield a similar result for the minimum acceleration time, (e.g. \citealt{kc15}):
\begin{equation}
\begin{split}
\label{eq:t_acc_rec}
t_{\rm acc}&= \frac{1}{\epsilon_0} \frac{r_L}{c}\\
& \sim 2\times10^{-4} \left(\frac{B}{100\ \rm G}\right)^{-3/2}\left(\frac{h \nu}{1\ \rm keV}\right)^{1/2}\left(\frac{\epsilon_0}{0.1}\right)^{-1} \ \mathrm{s}.
\end{split}
\end{equation} 
$\epsilon_0 = E/B_0 \leq 1$ is the reconnection rate, and simulations find an $\epsilon_0\sim 0.1$ (e.g. \citealt{kms13}). The fastest particles can be accelerated roughly the same whether the particles are accelerated via shocks or magnetic reconnection in a mildly relativistic outflow.

During a flare, $t_{\rm acc}$ is the smallest time we should expect the X-rays to lag the IR photons (assuming the IR photons are produced by electrons at the initial stage of acceleration). The lag time is so small we should expect the infrared and X-rays to be simultaneous. For example the X-ray lags on minute timescales (with respect to IR emission) reported by \cite{ys12}---who argue an inverse Compton origin for the X-ray flares---are unlikely to be explained by delayed particle acceleration and subsequent synchrotron emission in both bands. If further studies of the coupling between the IR/X-ray variability of~\sgra~indicated that the IR lags the X-ray, synchrotron emission would not be able to explain the observed time lag. This calculation does not, however, take into account synchrotron cooling of the electrons during the acceleration process, which is one of our predicted scenarios. Also we consider here only the time to accelerate particles, not the time delays we may expect for shocks or magnetic reconnection events to develop in a relativistic outflow, which is an interesting further point to explore in the context of \sgra's IR/X-ray variability. For example, as discussed in section \ref{sec:intro}, studies of \sgra's X-ray variability have allowed inferences regarding the particle acceleration process, with many finding magnetic reconnection to be a viable process, likely in a fast-cooling regime \citep{nei13,li15,dib16}---we note that this supports our sychrotron-dominated model scenario.\\
\indent Our new joint-fitting technique strongly favours a scenario in which the most sub-Eddington accreting black holes (quiescence down to $l_{X}\sim10^{-9}-10^{-8}$) have very compact jet bases, on the order of a few gravitational radii. This holds regardless of whether the emission is dominated by non-thermal synchrotron from a population of accelerated electrons (accelerated within a few $r_g$ of the black hole), or a high-density sub-equipartition flow which produces dominant SSC emission (or potentially a mixture of these two processes). We find that the particle acceleration component in the outer jet recedes, leaving evidence for another kind of weak acceleration in the inner accretion flow \citep{yqn03}, or an inverse Compton-dominated jet base. We echo the statements made by \cite{plot15} that further well-sampled SEDs of quiescent BHBs are required to draw more precise conclusions regarding the accretion/jet dynamics and local conditions, in particular breaking the degeneracy between synchrotron or SSC domination in such weakly accreting systems, and what is driving the change from one regime to another. In addition to this, the efforts of the Event Horizon Telescope (EHT) \citep{doe08} to probe the inner regions of~\sgra's accretion flow will shed light on local plasma conditions, and may reveal more about the plausibility of these model scenarios. A parallel strategy is to improve our modelling, and in the near future we will implement self-consistent, relativistic MHD flow solutions to reduce the free parameters in our jet modelling, in particular the geometrical quantities and their relationship to the internal properties (e.g. \citealt{pol10,pol13,pol14}). This will be presented in an upcoming paper (Ceccobello et al., in prep).\\
\indent In the future we hope to build further upon our results here based upon more recent broadband observations (X-ray/radio/optical-IR) of A0620 (e.g. \citealt{mbb15}) that may give more insight into the outflow structure of A0620. Such insights would come primarily from the radio spectrum of A0620, since in our modelling we are limited by the lack of radio coverage (with only the 8.5 GHz flux). 
 %%%% ACKNOWLEDGEMENTS %%%%%%%%%%%
 \section*{Acknowledgements}
This research has made use of \texttt{ISIS} functions provided by ECAP/Remeis observatory and MIT (http://www.sternwarte.uni-erlangen.de/\texttt{ISIS}/). \\
\indent RC is thankful for support from NOVA (Dutch Research School for Astronomy).  SM  and RC  are grateful to the University of Texas in Austin for its support through a Tinsley Centennial Visiting Professorship. RC thanks Rik van Lieshout for useful discussions on the use of Markov Chain Monte Carlo in parameter exploration. 
 %%%% BIBLIOGRAPHY %%%%%%%%%%%

%%%%% TEMPORARY LOCATION FOR FIGURES TABLES AT END OF DOCUMENT %%%%%%%%%%%%%%%%%%%%%%%%%%%%
 
%%%%% APPENDIX %%%%%%%%%%%%%%%
\appendix
\section[]{Broadband spectra of~\sgra~and A0620}
 %\floatplacement{table}{H} - NOTE TO PUBLISHER, THIS WAS ONLY NECESSARY IN THE VERSION WITH BOLD HIGHLIGHTED TEXT, JUST AN ISSUE WITH FLOAT PLACEMENT OF THE APPENDIX TABLES
\begin{table}
 \centering
 \caption{A0620 radio-to-FUV spectrum}
 \begin{tabular*}{0.47\textwidth}{@{\extracolsep{\fill}}lcr}
 \hline
 \hline
 $\nu$ (Hz) & $I_{\nu}$ (mJy) & Instrument \\
 \hline
 $8.50\times10^9$ & $0.051\pm0.007$ & VLA$^{\mathrm{a}}$\\
 $1.25\times10^{13}$ & $0.121\pm0.065$ & \textit{Sptizer}$^{\mathrm{a}}$ \\ 
 $3.75\times10^{13}$ & $0.305\pm0.031$ & \textit{Spitzer}$^{\mathrm{a}}$ \\ 
 $6.66\times10^{13}$ & $0.380\pm0.038$  & \textit{Spitzer}$^{\mathrm{a}}$ \\ 
 $1.40\times10^{14}$ & $1.275\pm0.188$ & ANDICAM (CTIO)$^{\mathrm{b}}$ \\ 
 $1.84\times10^{14}$  & $1.738\pm0.240$ & \textit{SMARTS}$^{\mathrm{a}}$ \\
 $1.84\times10^{14}$ & $1.443\pm0.120$ & ANDICAM (CTIO)$^{\mathrm{b}}$ \\
 $2.40\times10^{14}$ & $1.471\pm0.136$ & ANDICAM (CTIO)$^{\mathrm{b}}$ \\         $3.62\times10^{14}$ & $1.744\pm0.161$ & \textit{SMARTS}$^{\mathrm{a}}$ \\
 $3.62\times10^{14}$ & $1.184\pm0.065$ & ANDICAM (CTIO)$^{\mathrm{b}}$ \\ 
 $5.46\times10^{14}$ & $0.880\pm0.065$ & \textit{SMARTS}$^{\mathrm{a}}$ \\
 $5.46\times10^{14}$ & $0.718\pm0.059$ & ANDICAM (CTIO)$^{\mathrm{b}}$ \\
 $6.71\times10^{14}$ & $0.406\pm0.037$ & ANDICAM (CTIO)$^{\mathrm{b}}$ \\
 $8.56\times10^{14}$ & $0.192\pm0.003$ & UVOT (\textit{Swift})$^{\mathrm{b}}$ \\
 $1.01\times10^{15}$ & $0.159\pm0.080$ & STIS (\textit{HST})$^{\mathrm{b}}$ \\
 $1.11\times10^{15}$ & $0.127\pm0.063$ & STIS (\textit{HST})$^{\mathrm{b}}$ \\
 $1.20\times10^{15}$ & $0.114\pm0.057$ & STIS (\textit{HST})$^{\mathrm{b}}$ \\
 $1.30\times10^{15}$ & $0.132\pm0.066$ & STIS (\textit{HST})$^{\mathrm{b}}$ \\
 $1.47\times10^{15}$ & $0.108\pm0.054$ & STIS (\textit{HST})$^{\mathrm{b}}$ \\
 $1.78\times10^{15}$ & $0.017\pm0.007$ & COS (\textit{HST})$^{\mathrm{b}}$ \\
 $1.87\times10^{15}$ & $0.017\pm0.005$ & COS (\textit{HST})$^{\mathrm{b}}$ \\
 $2.04\times10^{15}$ & $0.022\pm0.002$ & COS (\textit{HST})$^{\mathrm{b}}$ \\
 $2.22\times10^{15}$ & $0.025\pm0.002$ & COS (\textit{HST})$^{\mathrm{b}}$ \\
 $2.34\times10^{15}$ & $0.022\pm0.006$ & COS (\textit{HST})$^{\mathrm{b}}$ \\
 $2.60\times10^{15}$ & $0.033\pm0.014$ & COS (\textit{HST})$^{\mathrm{b}}$ \\
 \hline
  \end{tabular*}
  \label{tab:A2}\\
 \textbf{Notes.} $^{\mathrm{a}}$ Data taken from \cite{gal06,gal07}. \\
 $^{\mathrm{b}}$ Data taken from \cite{fron11}
 \end{table}

\begin{table}
\centering
   \caption{\sgra radio-to-submm spectrum}
 \begin{tabular*}{0.47\textwidth}{@{\extracolsep{\fill}}lcr}
 \hline
 \hline
 $\nu$ (GHz) & $I_{\nu}$ (mJy) & Instrument \\
 \hline
 $0.33$ & $220\pm60$ & VLA$^{\mathrm{c}}$ \\
$0.64$ & $450\pm100$ & GMRT$^{\mathrm{a}}$ \\
$1.2$ & $520\pm90$ & VLA$^{\mathrm{a}}$ \\
$1.3$ & $520\pm60$ & VLA$^{\mathrm{a}}$ \\
$1.5$ & $620\pm60$ & VLA$^{\mathrm{a}}$ \\
$1.6$ & $592\pm28$ & VLA$^{\mathrm{d}}$ \\
$1.8$ & $630\pm50$ & VLA$^{\mathrm{a}}$ \\
$3.1$ & $702\pm32$ &VLA$^{\mathrm{d}}$ \\
$4.86$ & $660\pm40$ & VLA$^{\mathrm{a}}$ \\
$5.4$ & $870\pm118$ & VLA$^{\mathrm{d}}$ \\
$8.46$ & $690\pm30$ & VLA$^{\mathrm{a}}$ \\
$9$ & $932\pm129$ & VLA$^{\mathrm{d}}$ \\
$14$ & $1075\pm135$ & VLA$^{\mathrm{d}}$ \\
$14.9$ & $920\pm60$ & VLA$^{\mathrm{a}}$  \\
$21.1$ & $1164\pm52$ & VLA$^{\mathrm{d}}$ \\
$22.4$ & $1060\pm60$ & VLA$^{\mathrm{a}}$ \\
$32$ & $1382\pm87$ & VLA$^{\mathrm{d}}$ \\
$40.9$ & $1485\pm73$ & VLA$^{\mathrm{d}}$ \\
$43$ & $1600\pm200$ & VLA$^{\mathrm{c}}$ \\
$95$ & $2376\pm187$ & ALMA$^{\mathrm{e}}$ \\
$97$ & $2403\pm186$ & ALMA$^{\mathrm{e}}$ \\
$105$ & $2555\pm197$ & ALMA$^{\mathrm{e}}$ \\
 $107$ & $2597\pm231$ & ALMA$^{\mathrm{e}}$ \\
$216.8$ & $3677\pm762$ & SMA$^{\mathrm{d}}$ \\
$218$ & $3667\pm650$ & ALMA$^{\mathrm{d}}$ \\
$220$ & $3661\pm652$ & ALMA$^{\mathrm{d}}$ \\
$223.9$ & $3391\pm489$ & SMA$^{\mathrm{d}}$ \\
$230$ & $3300\pm300$ & JCMT$^{\mathrm{a}}$\\
$231.9$ & $3676\pm664$ & ALMA$^{\mathrm{d}}$ \\
$233.8$ & $3704\pm680$ & ALMA$^{\mathrm{d}}$ \\
$238.2$ & $3310\pm424$ & SMA$^{\mathrm{d}}$ \\
$266.8$ & $3369\pm96$ & SMA$^{\mathrm{d}}$ \\ 
$274$ & $3526\pm697$ & SMA$^{\mathrm{d}}$ \\
$331.1$ & $3205\pm1074$ & SMA$^{\mathrm{d}}$ \\
$338.3$ & $3436\pm863$ & SMA$^{\mathrm{d}}$ \\
$341.6$ & $3602\pm866$ & ALMA$^{\mathrm{d}}$ \\
$343.6$ & $3609\pm870$ & ALMA$^{\mathrm{d}}$ \\
$351.7$ & $3595\pm884$ & ALMA$^{\mathrm{d}}$ \\
$352.6$ & $4890\pm721$ & SMA$^{\mathrm{d}}$ \\
$353.6$ & $3553\pm860$ & ALMA$^{\mathrm{d}}$ \\
$375$ & $3500\pm500$ & JCMT$^{\mathrm{a}}$\\
$500$ & $4000\pm1200$ & JCMT$^{\mathrm{a}}$\\
$666$ & $3000\pm1000$ & JCMT$^{\mathrm{a}}$\\
$850$ & $7000\pm2000$ & CSO-JCMT$^{\mathrm{b}}$\\
\hline
\end{tabular*} \\
  \label{tab:A1}
 \textbf{Notes.} Data taken from \protect\cite{zyl95}$^{\mathrm{a}}$,
 \protect\cite{ser97}$^{\mathrm{b}}$, \protect\cite{an05}$^{\mathrm{c}}$,
  \protect\cite{bow15}$^{\mathrm{d}}$, and \protect\cite{bri15}$^{\mathrm{e}}$,
  shown from low-to-high frequency.
 \end{table}

\label{lastpage}
\end{document}